\documentclass[%
 reprint,
 nofootinbib,
 amsmath,amssymb,
 aps,
]{revtex4-2}

\usepackage{graphicx}
\usepackage{dcolumn}
\usepackage{bm}
\usepackage{mathtools}
\usepackage{mathrsfs}
\usepackage{amssymb}
\usepackage{siunitx}
\usepackage{hyperref}
\usepackage{enumerate}
\usepackage[usenames,dvipsnames]{xcolor}
\hypersetup{
    colorlinks=true,
    citecolor=BlueViolet,
    linkcolor=Mahogany,
    urlcolor=CadetBlue,
}
\newcommand{\defi}{\equiv}
\newcommand{\wt}{\widetilde}

\newcommand{\pd}{\partial}

\newcommand{\osc}{\text{osc}}
\newcommand{\vac}{\text{vac}}
\newcommand{\rM}{\mathrm{M}}
\newcommand{\re}{\mathrm{e}}
\newcommand{\rc}{\mathrm{c}}
\newcommand{\ro}{\mathrm{o}}
\newcommand{\Pl}{\text{pl}}
\newcommand{\trans}{\text{T}}
\newcommand{\crit}{\text{crit}}
\allowdisplaybreaks[4]

\begin{document}

\title{Observing axions through photon ring dimming of black holes}

\author{Kimihiro Nomura}
 \email{knomura@stu.kobe-u.ac.jp}

\author{Kaishu Saito}
 \email{184s151s@stu.kobe-u.ac.jp}

\author{Jiro Soda}
 \email{jiro@phys.sci.kobe-u.ac.jp}
 
\affiliation{%
Department of Physics, Kobe University, Kobe 657-8501, Japan
}%

\date{\today}

\begin{abstract}
    It is known that magnetic fields exist near black holes and photons can go around black holes due to strong gravity. Utilizing these facts, we can probe hypothetical pseudoscalar particles, so-called axions. In fact, photons can be converted into axions when they propagate in a magnetic field. The conversion of such photons into axions leads to a dimming of the photon ring around the black hole shadow. We show that  photon ring dimming can occur efficiently for supermassive black holes. Remarkably, it turns out that the maximal dimming rate of the photon ring is 25\%. In the case of M87$^*$, the dimming of 10\% will be observed in the X-ray and gamma-ray bands if the angular resolution of $10^{-5}\,\si{arcsec}$ is achieved.
    The frequency band and the magnitude of the dimming depend on the axion-photon coupling and axion mass. Hence, the distorted spectrum of the photon ring provides a novel tool for detecting axions.
\end{abstract}

\maketitle


\section{Introduction}
\label{sec:introduction}

Axions are hypothetical pseudoscalar particles originally introduced to solve the strong CP problem in quantum chromodynamics \cite{Peccei:1977hh, Weinberg:1977ma, Wilczek:1977pj, Kim:1979if, Shifman:1979if, Dine:1981rt, Zhitnitsky:1980tq}. 
Intriguingly, pseudoscalar particles will also arise ubiquitously in string theory~\cite{Svrcek:2006yi}. We refer to these pseudoscalar particles simply as axions. 
Axions can play important roles in cosmology \cite{Marsh:2015xka}. Indeed, heavy axions can realize slow-roll inflation naturally \cite{Freese:1990rb, Kim:2004rp, Dimopoulos:2005ac} because of shift symmetry.
Light axions can be dark matter~\cite{Preskill:1982cy, Abbott:1982af, Dine:1982ah, Hui:2016ltb, Chadha-Day:2021szb}.
Axions with the mass $10^{-33}~\si{eV}$ can mimic a cosmological constant \cite{Frieman:1995pm, Choi:1999xn, Copeland:2006wr}. 
Thus, it is worth probing axions and their mass from the cosmological point of view.

One of the crucial properties of axions is that they interact with photons through the coupling $\mathcal{L}_{\text{int}} = - (g_{a\gamma} / 4) \phi F_{\mu\nu} \wt{F}^{\mu\nu}$, where $g_{a\gamma}$ is the axion-photon coupling constant, $\phi$ is the axion field, $F_{\mu\nu}$ is the electromagnetic field strength, and $\wt{F}_{\mu\nu}$ is its dual. 
An interesting consequence of this interaction in the presence of a magnetic field is the conversion from photons into axions and vice versa~\cite{Maiani:1986md, Raffelt:1987im}. This photon-axion conversion phenomenon is the basic principle \cite{Sikivie:1983ip, Irastorza:2018dyq} to search for solar axions \cite{CAST:2017uph, Armengaud:2014gea} and axion dark matter \cite{ADMX:2009iij}. 
Photon-axion conversion has been widely discussed in cosmological and astrophysical contexts. For example, it is argued that high-energy photons from extragalactic sources propagate to us through the conversion from photons into axions and reconversion from axions into photons: 
Otherwise, such photons will be annihilated by electron-positron pair production~\cite{DeAngelis:2007dqd, Simet:2007sa, Sanchez-Conde:2009exi, Mirizzi:2009aj, Meyer:2013pny, Kohri:2017ljt}.
There are proposals to account for the recent detections of high-energy gamma-ray photons based on this idea~\cite{Zhang:2022zbm, Galanti:2022pbg, Troitsky:2022xso, Baktash:2022gnf, Lin:2022ocj, Gonzalez:2022opy, Nakagawa:2022wwm, Carenza:2022kjt, Galanti:2022xok}. 
It is also suggested that the conversion will lead to spectral distortions of the cosmic microwave background \cite{Yanagida:1987nf, Mirizzi:2009nq, Tashiro:2013yea} and X-/gamma-rays from high-energy sources such as active galactic nuclei \cite{Hooper:2007bq, Hochmuth:2007hk, DeAngelis:2007wiw, HESS:2013udx, Fermi-LAT:2016nkz, Marsh:2017yvc, Zhang:2018wpc, Reynolds:2019uqt}. On the other hand, axions could be produced in the cores of supernovae, super clusters, or white dwarfs, and they will be converted into photons that we may observe~\cite{Payez:2014xsa, Dessert:2020lil, Dessert:2021bkv}. Hence, in any case, the lack of observational signatures can be translated into a constraint on the coupling constant $g_{a\gamma}$. Together with observations \cite{Noordhuis:2022ljw, Dolan:2022kul, Dessert:2022yqq}, we obtain the upper bound of the couling constant as $g_{a\gamma} \lesssim 10^{-11}\text{--}10^{-10} ~ \si{GeV^{-1}}$ in the mass range  $m_a \lesssim 10^{-5}~\si{eV}$.

Recently, the Event Horizon Telescope observed a polarized synchrotron emission at $230 ~ \si{GHz}$ from near the event horizon of the black hole in the center of the M87 galaxy (M87*), and reported that the strength of the magnetic field is $1\text{--}30 ~\si{Gauss}$~\cite{EventHorizonTelescope:2021srq}. It is expected that magnetic fields of these orders of magnitude are commonly present in the vicinity of black holes in our universe. Hence, it is interesting to investigate photon-axion conversion around black holes. In this case, the propagation length of photons required for the conversion to axions is typically comparable to or longer than the horizon radius of supermassive black holes. Thus, one might think that a magnetic field maintained over the radial distance of that scale is necessary for conversion. However, we should note that the strong gravity of the black hole allows the photons to stay in its vicinity for a certain period of time. Specifically, a black hole spacetime has a photon sphere, on which unstable circular orbits of photons exist. For photons emitted from a source outside the black hole, a part of the photons will first approach the photon sphere, then stay around the sphere for a certain period of time, and finally escape from the sphere, that we observe. While orbiting the sphere, the photons stay at a nearly constant radius around the black hole. This fact automatically guarantees that the magnetic field is maintained during propagation. Previously, the conversion of photons near the photon sphere has been studied in Refs.~\cite{Saito:2021sgq,OuldElHadj:2021fqi} focusing on conversion to gravitons. For gravitons, coupling to photons is suppressed by the Planck scale, $M_\text{Pl}^{-1} \sim 10^{-19} ~ \si{GeV}^{-1}$. On the other hand, coupling of photons to axions is less constrained by observations.
Thus, photon-axion conversion could be more effective.
Therefore, in this paper, we investigate photon-axion conversion near the photon sphere of black holes.

Photon-axion conversion near the photon sphere is quite relevant to observations of the near black hole region, in particular, the bright ring-like image (``photon ring'') created around the dark region (``black hole shadow''), which is observed by the Event Horizon Telescope.
In fact, the number of photons emitted from the vicinity of a black hole will be reduced, i.e., \textit{the photon ring will be darkened} by conversion into axions. Remarkably, we will see that photon ring dimming can occur efficiently for supermassive black holes, and, in the case of M87$^*$, the dimming will be 10\% in the X-ray and gamma-ray bands for the axion with coupling $g_{a\gamma} \sim 10^{-11}~\si{GeV}^{-1}$ and mass $m_a \lesssim 10^{-7} ~ \si{eV}$. It will be shown that, in general, the frequency band and magnitude of the dimming depend on $g_{a\gamma}$ and $m_a$. Hence, observing the distorted spectrum of the photon ring provides a novel tool for probing the properties of axions.

This paper is organized as follows. In Sec.~\ref{sec:p-a_conversion}, we briefly review photon-axion conversion in a magnetic field and clarify the parameter region in which conversion efficiently occurs. In Sec.~\ref{sec:photon_sphere}, we study conversion near the photon sphere of black holes. Section \ref{sec:conclusion} is devoted to the conclusion. Appendices provide several supplements for the main sections.

We set $c = \hbar = k_\mathrm{B} = G = 1$, where $c$ is the speed of light, $\hbar$ is the reduced Planck constant, $k_\mathrm{B}$ is the Boltzmann constant, and $G$ is the Newton constant.
For electromagnetism, Gaussian units commonly used in astrophysics are applied in the main sections, while rationalized Heaviside--Lorentz units are used in Appendix \ref{app:conversion}. 

\section{Photon-axion Conversion}
\label{sec:p-a_conversion}

In this section, we briefly review the photon-axion conversion phenomenon in an external magnetic field.
We will see that conversion can efficiently occur for X-ray and gamma-ray propagating in the vicinity of black holes such as M87$^*$.

\subsection{Conversion probability}
\label{subsec:conversion_probability}

We consider photons propagating in an external magnetic field.
The photons with polarization parallel to the magnetic field are converted into axions.
Let us list parameters relevant to the conversion:
\begin{itemize}
    \item $\omega$: frequency of the propagating photons,
    \item $B$: magnetic field perpendicular to the photon propagation,
    \item $m_a$: axion mass,
    \item $g_{a\gamma}$: axion-photon coupling constant,
    \item $n_e$: number density of the electron in the medium.
\end{itemize}
The number density of the electron $n_e$ is used to determine the plasma frequency,
\begin{align}
    \omega_{\Pl} &\defi \sqrt{\frac{4\pi \alpha n_e}{m_e}}
    \notag \\
    &= 3.7 \times 10^{-11} ~ \si{eV} \sqrt{ \frac{n_e}{\si{cm}^{-3}} },
\end{align}
where $m_e = 511 \, \si{keV}$ is the electron mass, and $\alpha = 1/137$ is the fine-structure constant.

After the photons propagate over a distance $z$, the probability of conversion from photons to axions is given by (see Appendix \ref{app:conversion} or Refs.~\cite{Raffelt:1987im, Hochmuth:2007hk, Masaki:2017aea})
\begin{align}
    P_{\gamma \to a}(z) = \left( \frac{\Delta_\rM}{\Delta_\osc / 2} \right)^2 \sin^2 \left( \frac{\Delta_\osc}{2} z \right),
    \label{probability}
    \\
    \Delta_{\osc}^2 \defi (\Delta_{\Pl} - \Delta_{\vac} - \Delta_a)^2 + 4 \Delta_\rM^2.
    \label{osclength}
\end{align}
Here, $\Delta_\rM$, $\Delta_a$, $\Delta_{\Pl}$, and $\Delta_\vac$ are given, respectively, by
\begin{align}
    \Delta_\rM 
    &= 9.8 \times 10^{-23} ~\si{eV} \left( \frac{g_{a\gamma}}{10^{-11} \, \si{GeV}^{-1}} \right) \left( \frac{B}{\si{Gauss}} \right),
    \label{eq_Delta_M}
    \\
    \Delta_a
    &= 5 \times 10^{-22} ~ \si{eV} \left( \frac{m_a}{\si{n eV}} \right)^2 \left( \frac{\si{keV}}{\omega} \right),
    \label{eq_Delta_a}
    \\
    \Delta_{\Pl}
    &= 6.9 \times 10^{-25} ~ \si{eV} \left( \frac{n_e}{\si{cm}^{-3}} \right) \left( \frac{\si{keV}}{\omega} \right),
    \label{eq_Delta_pl}
    \\
    \Delta_{\vac} 
    &= 9.3 \times 10^{-29} ~ \si{eV} \left( \frac{\omega}{\si{keV}} \right) \left(  \frac{B}{\si{Gauss}}\right)^2.
    \label{eq_Delta_vac}
\end{align}
The $\Delta_\rM$ determined by $g_{a\gamma}$ and $B$ is an essential parameter for conversion. 
The effect of a finite axion mass $\Delta_a$, plasma oscillations $\Delta_{\Pl}$, and the Euler--Heisenberg effective Lagrangian in an external magnetic field incorporating the one-loop corrections of electrons $\Delta_{\vac}$, generically suppress the conversion.
Note that, for the validity of the present framework, at least the following three conditions should be satisfied;
\begin{itemize}
\item[1)] $(\alpha / (45 \pi)) (B / B_\text{crit})^2 \ll 1$ where $B_{\text{crit}} \defi m_e^2/ \sqrt{4\pi \alpha} = 4 \times 10^{13} ~ \si{Gauss}$, which comes from the validity of the Euler--Heisenberg effective Lagrangian, 
\item[2)] $\omega \gg m_a$, i.e., axions should be relativistic, 
\item[3)] $\omega \gg \omega_{\Pl}$ so that photons can propagate in a surrounding plasma.
\end{itemize}

\subsection{Efficient conversion}
\label{subsec:efficient_conversion}

The most efficient conversion can be realized when $(\Delta_\Pl - \Delta_\vac - \Delta_a)^2 \ll 4 \Delta_\rM^2$ so that $\Delta_\osc \simeq 2 \Delta_\rM$, i.e., the prefactor of the probability \eqref{probability} approaches unity.
In this case, the typical length scale of the conversion reads 
\begin{align}
    \Delta_\osc^{-1} &\simeq (2 \Delta_\rM)^{-1} 
    \notag \\
    &= 1.0 \times 10^{17} ~ \si{cm} \left( \frac{\si{Gauss}}{B} \right) \left( \frac{10^{-11} ~ \si{GeV}^{-1}}{g_{a\gamma}} \right)
    \notag \\
    &= 3.4 \times 10^2 \times (2 \times 10^9 M_\odot) 
    \notag \\
    &\quad \times \left( \frac{\si{Gauss}}{B} \right) \left( \frac{10^{-11} ~ \si{GeV}^{-1}}{g_{a\gamma}} \right).
    \label{convlength}
\end{align}
We can see that, for $B \sim 10^{1 \text{--} 2} ~ \si{Gauss}$, the conversion length will be
comparable to the Schwarzschild radius of a supermassive black hole with mass $\sim 10^{9 \text{--} 10} M_\odot$, if $g_{a\gamma} \sim 10^{-11} ~ \si{GeV}^{-1}$.
In fact, observations of M87$^*$ tell us that it has a mass of $6 \times 10^9 M_\odot$ \cite{EventHorizonTelescope:2019dse} and a magnetic field around $30 ~ \si{Gauss}$ in the vicinity of the black hole \cite{EventHorizonTelescope:2021srq}.
Since photons can stay around the photon sphere of black holes for some period of time, we can expect that the conversion to axions efficiently occurs.

The above condition $(\Delta_\Pl - \Delta_\vac - \Delta_a)^2 \ll 4 \Delta_\rM^2$ is satisfied at least when the photon-axion mixing effect $\Delta_\rM$ dominates over the others $\Delta_a$, $\Delta_\Pl$, and $\Delta_\vac$.
The inequalities $\Delta_a \ll \Delta_\rM$, $\Delta_\Pl \ll \Delta_\rM$, and $\Delta_\vac \ll \Delta_\rM$ are respectively rewritten as
\begin{align}
    &5.1 \left( \frac{m_a}{\si{neV}} \right)^2
    \ll \left( \frac{g_{a\gamma}}{10^{-11} \, \si{GeV}^{-1}} \right) \left( \frac{\omega}{\si{keV}} \right) \left( \frac{B}{\si{Gauss}} \right),
    \label{efficient2}\\
    &7.0 \times 10^{-3} \left( \frac{n_e}{\si{cm}^{-3}} \right) 
    \notag \\
    &\quad \ll     
     \left( \frac{g_{a\gamma}}{10^{-11} \, \si{GeV}^{-1}} \right) \left( \frac{\omega}{\si{keV}} \right) \left( \frac{B}{\si{Gauss}} \right),
    \label{efficient1}
\end{align}
and
\begin{align}
    \left( \frac{\omega}{\si{keV}} \right) \left(  \frac{B}{\si{Gauss}}\right)
    \ll 
    1.1 \times 10^{6} \left( \frac{g_{a\gamma}}{10^{-11} \, \si{GeV}^{-1}} \right) .
    \label{efficient3}
\end{align}
Except for specific cases i) and ii) mentioned later, either Eq.~\eqref{efficient2} or Eq.~\eqref{efficient1} determines the lower bound of $\omega$ where conversion occurs efficiently.
On the other hand, the upper bound of $\omega$ for efficient conversion is given by Eq.~\eqref{efficient3}.

Even when the plasma effect is sizable, $\Delta_\Pl \gtrsim \Delta_\rM$, efficient conversion can be realized if $\Delta_\Pl$ is canceled by $\Delta_a$ or $\Delta_\vac$.
This is possible because $\Delta_\Pl$ contributes to $\Delta_\osc$ with opposite sign relative to $\Delta_a$ and $\Delta_\vac$\footnote{The definition of $\Delta_i$'s $(i=\rM, a, \Pl, \vac)$ here is not exactly the same as the definition in Ref.~\cite{Raffelt:1987im}. In our paper, $\Delta_i$'s are defined to be all positive. On the other hand, in Ref.~\cite{Raffelt:1987im}, they are defined so that the refractive index $n_i$ is expressed as $n_i = 1 + \Delta_i / \omega$. In particular, $\Delta_a$ and $\Delta_\Pl$ (denoted by $\Delta^{\text{gas}}$ in Ref.~\cite{Raffelt:1987im}) are opposite in sign.} as we can see in Eq.~\eqref{osclength}. Let us see these possibilities below:
\begin{enumerate}[i)]
    \item $\Delta_{\Pl} \simeq \Delta_a$: this resonance condition is equivalent to $m_a^2 \simeq \omega_\Pl^2$, i.e., 
    \begin{align}
        \left( \frac{m_a}{\si{n eV}} \right)^2 \simeq 1.4 \times 10^{-3}  \left( \frac{n_e}{\si{cm}^{-3}} \right).
        \label{resonance1-1}
    \end{align}
    Here we assumed that the sizable $\Delta_\Pl$ is compensated by the similarly sizable $\Delta_a$ which is much larger than $\Delta_\vac$, i.e., 
    \begin{align}
        \left( \frac{\omega}{\si{keV}} \right)^2 \left(\frac{B}{\si{Gauss}}\right)^2
        \ll 
        5.4 \times 10^{6} \left( \frac{m_a}{\si{neV}} \right)^2 .
        \label{resonance1-2}
    \end{align}
    Note that, in this case, the conversion can occur in all frequencies satisfying Eq.~\eqref{resonance1-2} (and the conditions for the present treatment to be justified).
    
    \item $\Delta_{\Pl} \simeq \Delta_\vac$: this is written as 
    \begin{align}
        \left( \frac{\omega}{\si{keV}} \right)^2 \left(  \frac{B}{\si{Gauss}}\right)^2 
        \simeq 
        7.4 \times 10^{3} \left( \frac{n_e}{\si{cm}^{-3}} \right) ,
        \label{resonance2-1}
    \end{align}
    which determines the resonance frequency.
    Here we assumed $\Delta_\vac \gg \Delta_a$, i.e,
    \begin{align}
        \left( \frac{\omega}{\si{keV}} \right)^2 \left(\frac{B}{\si{Gauss}}\right)^2 
        \gg 
        5.4 \times 10^{6} \left( \frac{m_a}{\si{neV}} \right)^2 .
        \label{resonance2-2}
    \end{align}    
\end{enumerate}

\subsection{Conversion probability in the $\omega$-$n_e$ plane}
\label{subsec:omega-n_e_plane}

\begin{figure}[htbp]
    \begin{minipage}[htb]{\hsize}
        \includegraphics[width=\hsize]{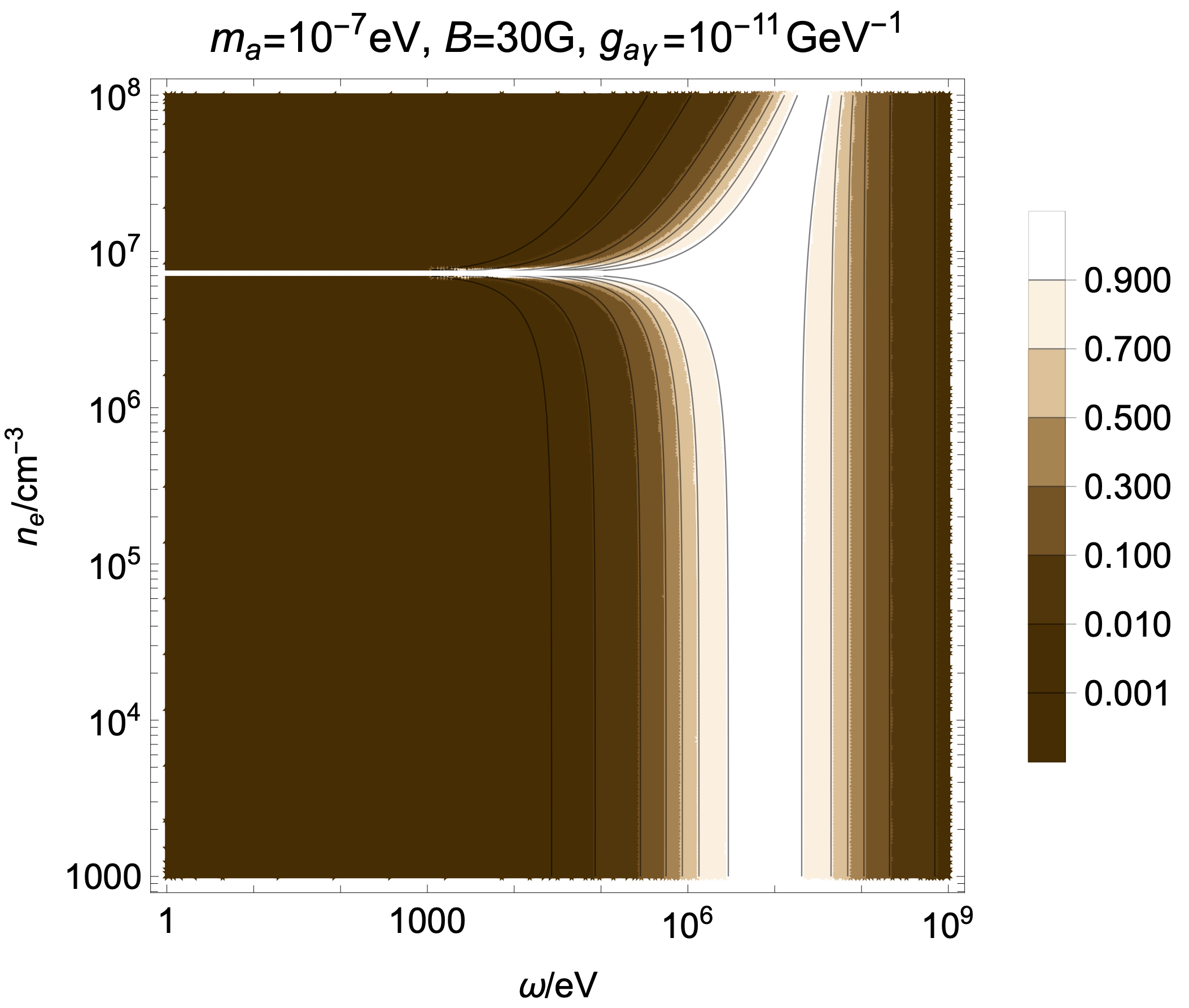}\\
        \includegraphics[width=\hsize]{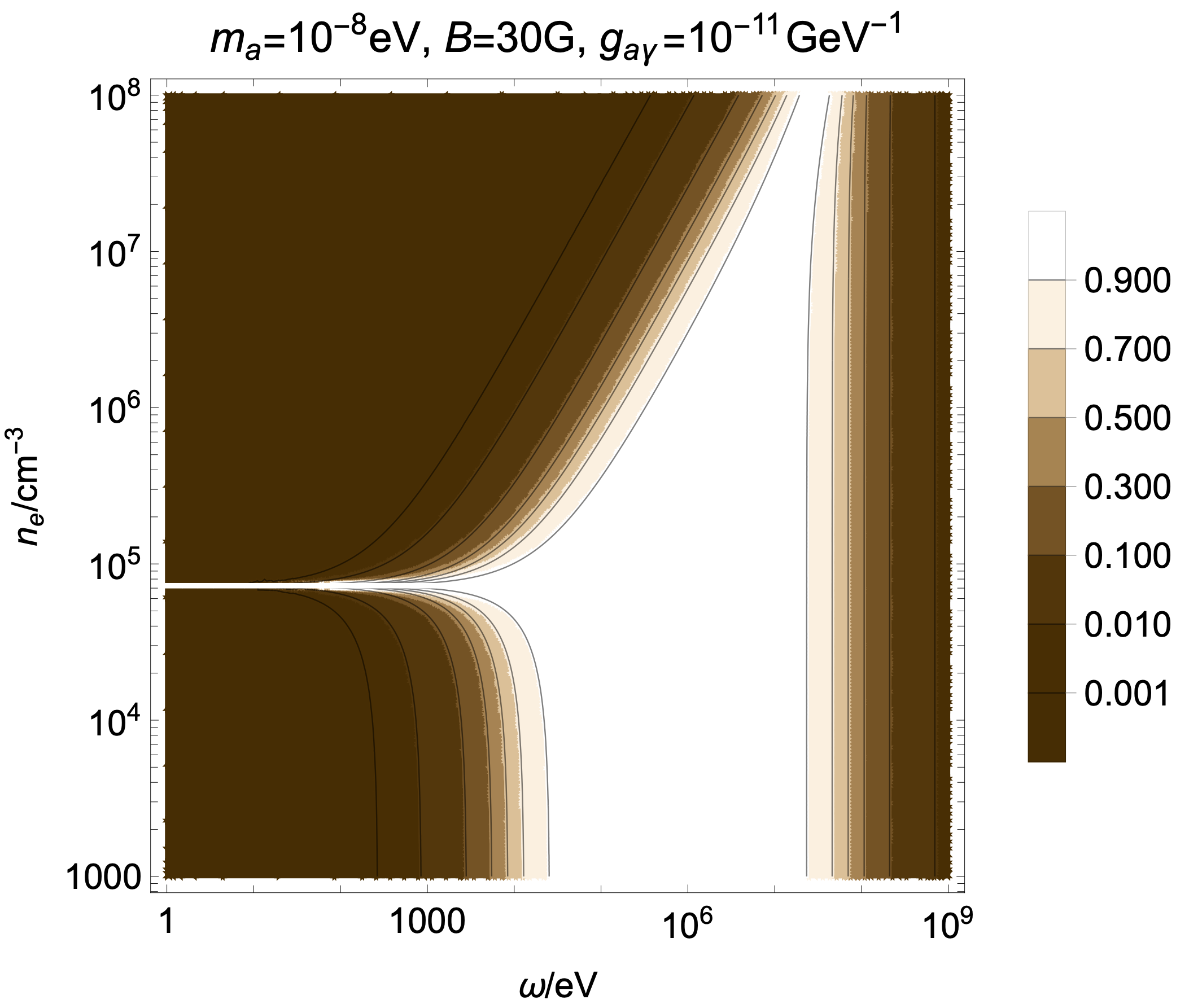}\\
        \includegraphics[width=\hsize]{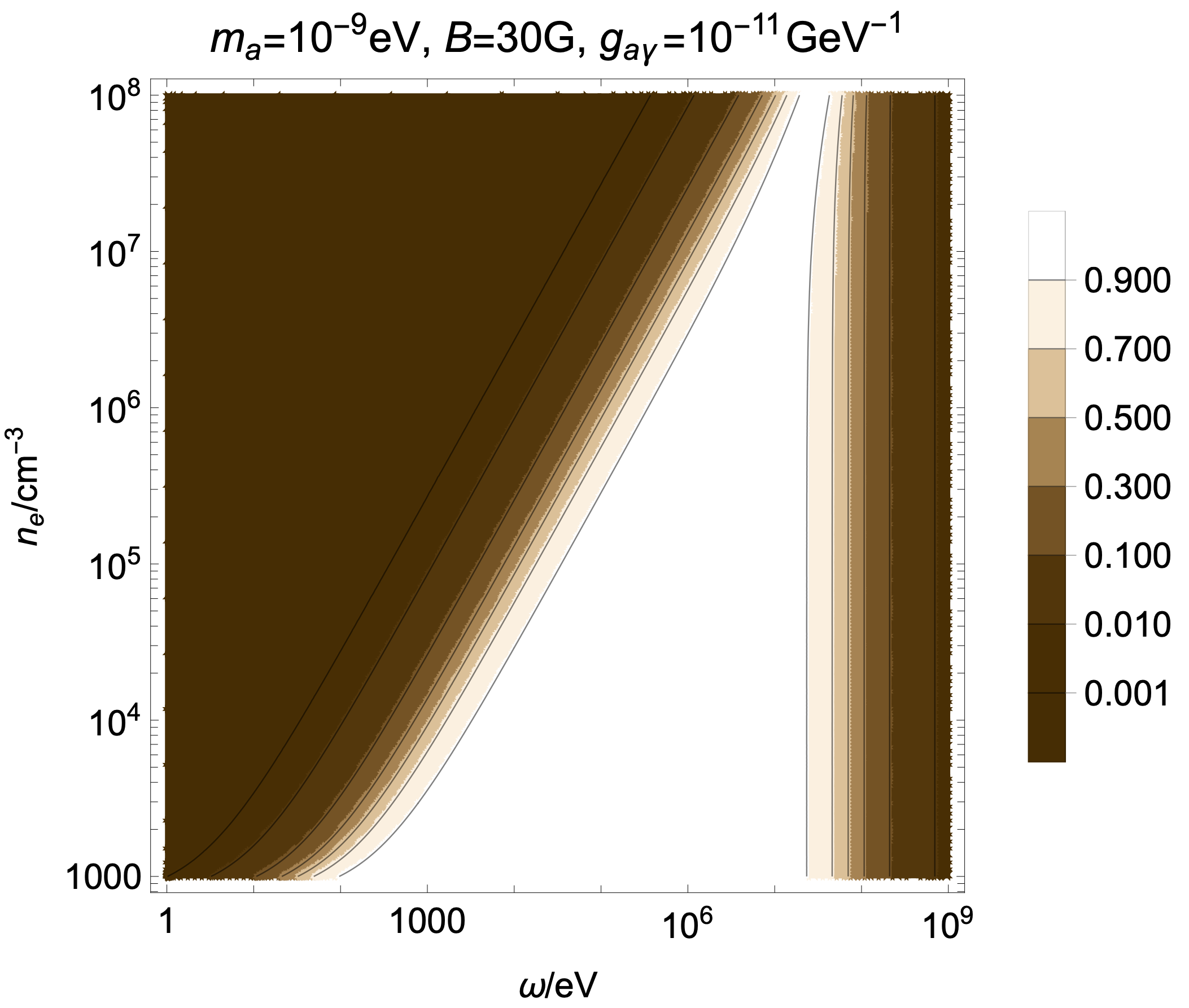}\\
    \end{minipage}
    \caption{Conversion probability omitting the distance-dependent oscillation factor, $[\Delta_\rM / (\Delta_{\osc}/ 2)]^2$, in the $\omega$-$n_e$ plane. We fix $B = 30 ~\si{Gauss}$ and $g_{a\gamma} = 10^{-11} ~ \si{GeV}^{-1}$. The axion mass is chosen as $m_a = 10^{-7}, 10^{-8}, 10^{-9}~\si{eV}$ from top to bottom.}
    \label{fig:omega-ne-3}
\end{figure}

In this subsection, we try to visualize the conversion probability in the parameter space to see its behavior at a glance.
To this end, we fix the axion-photon coupling to be $g_{a\gamma} = 10^{-11} ~ \si{GeV}^{-1}$.
From Event Horizon Telescope observations, we know the electron number density $\sim 10^{4 \text{--}7} ~ \si{cm^{-3}}$ and magnetic field $\sim 1$--$30 ~ \si{Gauss}$  near M87$^*$ \cite{EventHorizonTelescope:2021srq}.
Thus we consider $n_e$ around that range, and use $B = 30 ~ \si{Gauss}$ as a reference value.
The conversion probabilities omitting the $z$-dependent oscillation factor, i.e., $[\Delta_\rM / (\Delta_{\osc}/ 2)]^2$, in the $\omega$-$n_e$ plane for axion mass $m_a = 10^{-7}, 10^{-8}, 10^{-9}$ eV are shown in Fig.~\ref{fig:omega-ne-3}.
There, the region in which the conversion probability can approach unity is displayed in white.
The horizontal white lines extended toward the low-$\omega$ region correspond to case i) $\Delta_\Pl \simeq \Delta_a$.
We can also see that white bands are extended but getting narrower toward the upper-right, which will be connected to the line of resonance ii) $\Delta_\Pl \simeq \Delta_\vac$.

First, let us see the case  $m_a = 10^{-7}$ eV depicted in the top panel in Fig.~\ref{fig:omega-ne-3}.
Above the horizontal white line at $n_e = 7.3 \times 10^6 ~ \si{cm^{-3}}$, Eq.~\eqref{efficient1} determines the lower bound of $\omega$ for efficient conversion. Below that line, Eq.~\eqref{efficient2} gives the lower bound.
The upper bound of $\omega$ for efficient conversion is given by Eq.~\eqref{efficient3}. The white region  almost lies in $\omega \sim 10^{6\text{--}7}~\si{eV}$. For photons around these frequencies, $e^-$ - $e^+$ pair creations would be relevant, which makes discussion of photon-axion conversion subtle. 
If the axion mass is heavier than $10^{-7}~\si{eV}$, the window of the efficient conversion (white region) becomes narrower in the $\omega$-direction, and eventually closes except for the resonance lines due to i) and ii). This is because the inequalities \eqref{efficient2}--\eqref{efficient3} are hardly satisfied at the same time. 

In the case of a smaller axion mass, $m_a = 10^{-8}$ eV, the effect of $\Delta_a$ proportional to $m_a^2$ becomes smaller, and hence the white line corresponding to case i) moves toward the smaller $n_e$ as we can see in the middle panel in Fig.~\ref{fig:omega-ne-3}. The plasma effect represented by Eq.~\eqref{efficient1} determines the lower bound of $\omega$ for efficient conversion in a broad region.
Below the resonance line, Eq.~\eqref{efficient2} gives the lower bound.
The upper bound of $\omega$ for efficient conversion is again given by Eq.~\eqref{efficient3}. 

If the axion mass is $m_a = 10^{-9}$ eV, the plasma effect $\Delta_\Pl$ exceeds over the axion mass effect $\Delta_a$ in the whole space with $n_e \gtrsim 10^3 ~\si{cm^{-3}}$.
Thus, as we can see in the bottom panel in Fig.~\ref{fig:omega-ne-3}, Eq.~\eqref{efficient1} completely determines the lower bound of $\omega$ for efficient conversion in that space.
Still, the upper bound of $\omega$ for efficient conversion is determined by Eq.~\eqref{efficient3}. 

From Fig.~\ref{fig:omega-ne-3}, we can see that a broad white region exists for $m_a \lesssim 10^{-8} ~ \si{eV}$. This fact indicates that conversion will occur even in inhomogeneous plasma and magnetic fields, which are the cases in realistic situations.

\section{Conversion near photon spheres}
\label{sec:photon_sphere}

\subsection{Photon ring dimming}
\label{subsec:photon_ring_dimming}

We have seen that photons propagating over a distance $\sim \Delta_\osc^{-1}$ can be efficiently converted into axions if $\Delta_\osc \simeq 2\Delta_\rM$. The length scale $\Delta_\osc^{-1}$ is typically as long as or longer than the Schwarzschild radius of astrophysical black holes, such as M87$^*$. 
For photons propagating in the radial direction from a black hole, a strong magnetic field has to be maintained over the distance $\Delta_\osc^{-1}$ for conversion.
On the other hand, a black hole spacetime has a photon sphere, where unstable circular orbits for photons (or relativistic particles in general) exist.
For photons orbiting around the photon sphere, it is automatically guaranteed that the magnetic field is maintained during propagation since they stay at a nearly constant radius.
This fact indicates that conversion from photons to axions efficiently occurs around the photon sphere.

For simplicity, we take the Schwarzschild black hole spacetime,
\begin{align}
    ds^2 &= -f(r) dt^2 + \frac{1}{f(r)} dr^2 + r^2 (d\theta^2 + \sin^2 \theta \, d\phi^2),
    \label{sch_metric}
\end{align}
where we defined 
\begin{align}
    f(r) &= 1-\frac{2M}{r}
\end{align}
with $M$ being the mass of the black hole.
In this case, the photon sphere is located at the radius $r = 3M$. It is convenient to define the impact parameter $b$ of a particle as $b \defi L/E$, where $L$ is the conserved angular momentum along the geodesic, and $E$ is the conserved energy (see Appendix \ref{app:orbiting_time}). Photons with the critical impact parameter $b_\crit \defi 3 \sqrt{3} M$ can keep orbiting on the photon sphere unless disturbed by perturbations. 
For photons emitted from a source far outside the black hole with impact parameter $b$ slightly larger than the critical one $b_\crit$, the trajectories will first approach the photon sphere, then stay around the sphere for some period of time, and finally escape away from the sphere. We observe the photons survived against conversion. 

Let $d^3 N / dtd\omega_\rc db$ be the number of photons approaching the photon sphere with impact parameter $b$ close to $b_\crit$, per unit time $t$, unit frequency $\omega_\rc$, and unit impact parameter $b$.
Here, the subscript ``c'' of $\omega_\rc$ reminds us that it is measured in a local inertial frame at the photon sphere.
The photons with $b$ close to $b_\text{crit}$ stay orbiting in a region near the photon sphere, $3M < r \lesssim (3+\epsilon)M$ with a small $\epsilon (> 0)$.
The time staying there is given by $T(b) = -3\sqrt{3} M \ln |2( b-b_\crit ) / (\sqrt{3} \epsilon^2 M)|$ as shown in Eq.~\eqref{B17}. In terms of the proper distance, such photons travel for $z \simeq \sqrt{f(3M)}\, T(b) = T(b) / \sqrt{3}$. 
Thus, the number of photons converted into axions in that region per unit time $t$ and unit frequency $\omega_\rc$ is given by 
\begin{align}
    \frac{d^2N_{\gamma \to a}}{dt d\omega_\rc}
    &= \int_{b_{\crit}}^{b_{\crit}+\sqrt{3} \epsilon^2 M/2} 
    \!\!\! db \, \frac{1}{2} \left(\frac{d^3N}{dtd\omega_\rc db}\right) 
    P_{\gamma \to a} \left( \frac{T(b)}{\sqrt{3}} \right), 
    \label{axion_rate_0-1}
\end{align}
where 
\begin{align}
    P_{\gamma \to a}  \left( \frac{T(b)}{\sqrt{3}} \right)
    &= \left( \frac{\Delta_\rM}{\Delta_\osc / 2} \right)^2
    \notag \\
    &\quad \times  \sin^2 \left( - \frac{3M\Delta_\osc }{2} \ln \frac{2(b-b_\crit)}{\sqrt{3} \, \epsilon^2 M} \right).
    \label{axion_rate_0-2}
\end{align}
The factor $1/2$ in Eq.~\eqref{axion_rate_0-1} comes from the fact that only photons with polarization parallel to the external magnetic field can be converted into axions.
Taking the integration interval as $(b_{\crit}, b_{\crit} + \sqrt{3} \epsilon^2 M / 2)$ in Eq.~\eqref{axion_rate_0-1}, we can sum up photons which enter the region $3M < r \lesssim (3+\epsilon) M$ and escape out to infinity.
Note that $\Delta_\osc$ depends on the frequency of the photons $\omega_\rc$, and $B$ in $\Delta_\rM$ and $\Delta_\vac$ is a component of the magnetic field normal to the photon sphere.

Assuming that $d^3N/dt d\omega_\rc db$ in the integrand of Eq.~\eqref{axion_rate_0-1} does not vary significantly with respect to $b$, we can replace it by the value at $b= b_{\crit}$ as an approximation. 
Then, Eq.~\eqref{axion_rate_0-1} is recast as  
\begin{align}
    \frac{d^2N_{\gamma \to a}}{dt d\omega_\rc}
    &\simeq  \frac{1}{2} \left. \frac{d^3N}{dtd\omega_\rc db} \right|_{b = b_\crit}
    \notag \\
    &\quad \times \int_{b_\crit}^{b_\crit + \sqrt{3} \epsilon^2 M /2} db \, 
    P_{\gamma \to a} \left( \frac{T(b)}{\sqrt{3}} \right)
    \notag \\
    &= \frac{1}{2}\left. \frac{d^3N}{dtd\omega_\rc db} \right|_{b = b_\crit}
    \notag \\
    &\quad \times
    \left( \frac{\Delta_\rM}{\Delta_\osc / 2} \right)^2
    \frac{\sqrt{3} \, \epsilon^2 M}{4} \frac{(3M \Delta_\osc)^2}{1 + (3M\Delta_\osc)^2}.
    \label{axion_rate_0-3}
\end{align}
Consequently, the fraction of photons entering the region near the photon sphere that are converted into axions is 
\begin{align}
    \frac{d^2N_{\gamma \to a}}{dt d\omega_\rc} \bigg/ \frac{d^2N}{dt d\omega_\rc}
    \simeq \frac{1}{4} \left( \frac{\Delta_\rM}{\Delta_\osc / 2} \right)^2 \frac{(3M \Delta_\osc)^2}{1 + (3M\Delta_\osc)^2},
    \label{axion_rate_0-4}
\end{align}
which depends on $\omega_\rc$.
When we observe the vicinity of a black hole, a bright image like a ring (``photon ring'') can be seen around a dark region (``shadow''), which is created by photons traveling around the photon sphere.
The analysis here indicates that we will observe a dimming of the photon ring at the fraction \eqref{axion_rate_0-4} due to photon-axion conversion.
The above calculation is based on the assumption that photons propagate without scattering by surrounding plasma. In fact, the result is reliable when the mean free path of photons is sufficiently longer than $3M$, which is typically the case as shown in Appendix \ref{app:scattering}.

\begin{figure}
    \includegraphics[width=\hsize]{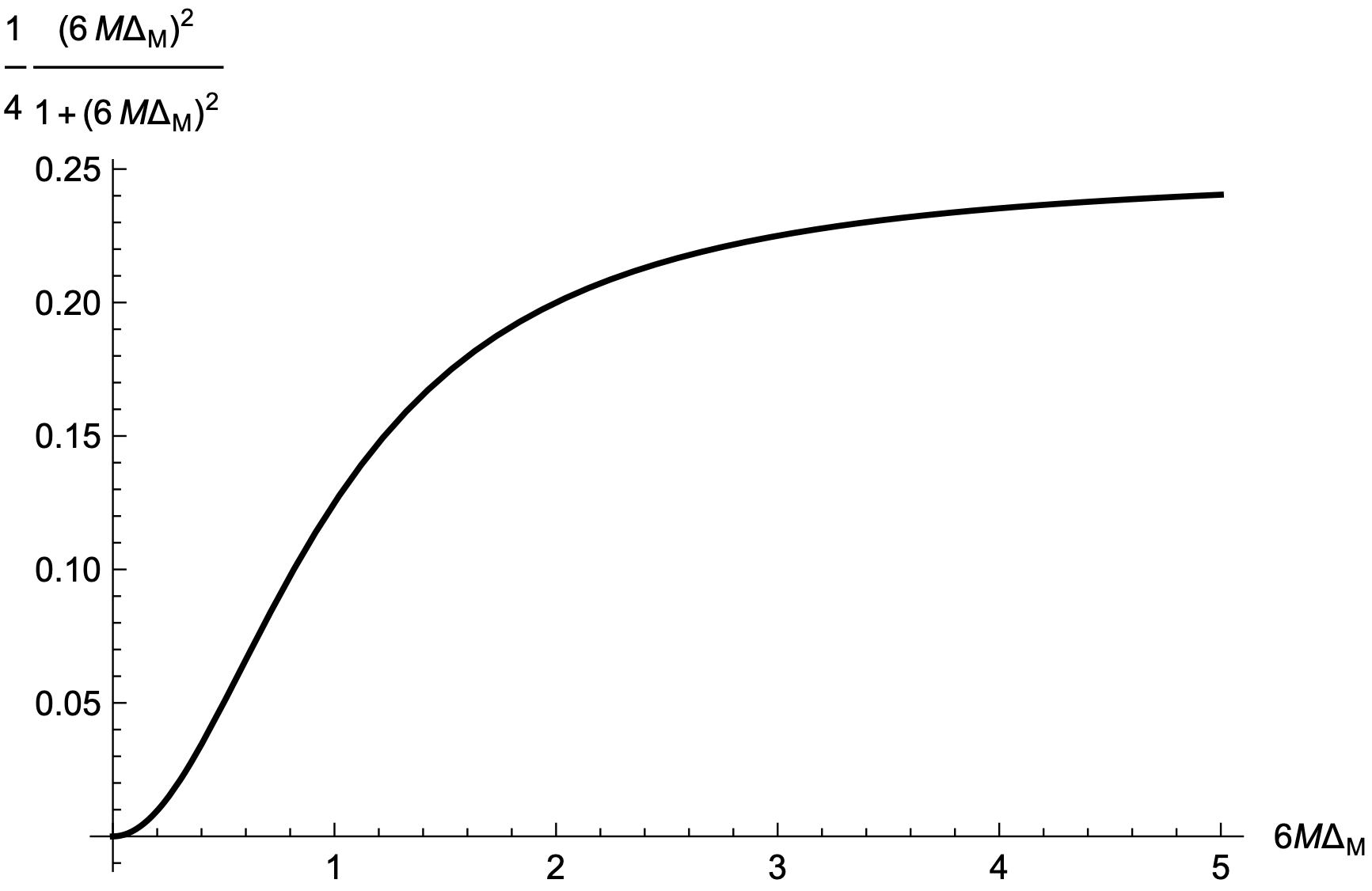}
    \caption{
    The dimming rate of a photon ring due to photon-axion conversion (in other words, the fraction of photons converted into axions) at the efficient case satisfying $\Delta_\osc \simeq 2 \Delta_\rM$ is plotted.}
    \label{fig:dimming}
\end{figure}

In particular, let us focus on the case of efficient conversion satisfying $\Delta_\osc /2 \simeq \Delta_\rM$ studied in Sec.~\ref{subsec:efficient_conversion}.
In this case, the fraction of photons converted into axions is given by
\begin{align}
    \frac{d^2N_{\gamma \to a}}{dt d\omega_\rc} \bigg/ \frac{d^2N}{dt d\omega_\rc}
    \simeq  
    \frac{1}{4} \frac{(6M \Delta_\rM)^2}{1 + (6M \Delta_\rM)^2}
    \notag \\
    (\text{if} ~ \Delta_\osc /2 \simeq \Delta_\rM).
    \label{axion_rate_0-4-2}
\end{align}
The key quantity $6M \Delta_\rM = 3M / (2 \Delta_\rM)^{-1}$ is the ratio of the photon sphere radius $3M$ to the conversion length $\Delta_\osc^{-1} \simeq (2 \Delta_\rM)^{-1}$, which reads 
\begin{align}
    6M \Delta_\rM 
    &= 4.4 \times 10^{-3} \left( \frac{M}{10^9 M_\odot} \right) \left( \frac{B}{\text{Gauss}} \right)
    \notag\\ 
    &\quad\times  \left( \frac{g_{a\gamma}}{10^{-11} ~\si{GeV}^{-1}} \right).
    \label{axion_rate_0-5}
\end{align}
If this value is much larger than unity, the magnitude of the dimming of a photon ring approaches 25\%. This is accounted for by the fact that only photons with polarization parallel to the magnetic field are converted, and photons with that polarization and axions are equally produced due to large mixing.
On the other hand, if $(6M \Delta_\rM)^2 \ll 1$, the magnitude of the dimming is approximated as $(6M \Delta_\rM)^2 \times$ 25\%. 
The dimming at efficient conversion as a function of $6M \Delta_\rM$ is plotted in Fig.~\ref{fig:dimming}. Note that $6M\Delta_\rM$ is proportional to the black hole mass $M$. Thus, supermassive black holes such as M87$^*$ are good candidates for 
strong dimming of the photon sphere. In contrast, for stellar mass black holes, it would be difficult to observe dimming unless there are strong magnetic fields compensating for the effect of the small black hole masses.

As for the supermassive black hole at the center of the galaxy M87, the mass and magnetic field around $r \sim 5 M$ are estimated as $M \simeq 6.2 \times 10^9 M_\odot$ \cite{EventHorizonTelescope:2019dse} and $B \sim 1$--$30 ~ \si{Gauss}$ \cite{EventHorizonTelescope:2021srq}, respectively.
The current observational constraint on axion-photon coupling is given by $g_{a\gamma} \lesssim 10^{-11}\text{--}10^{-10} ~ \si{GeV^{-1}}$ for $m_a \lesssim 10^{-5} ~\si{eV}$ as mentioned in the introduction. 
Thus, let us take $M = 6.2 \times 10^9 M_\odot$, $B = 30 ~ \si{Gauss}$, and $g_{a\gamma} = 10^{-11} ~ \si{GeV}^{-1}$ as a trial.
Then we have $6 M \Delta_\rM = 0.82$ and thus the dimming of the photon ring is
\begin{align}
    \frac{d^2N_{\gamma \to a}}{dt d\omega_\rc} \bigg/ \frac{d^2N}{dt d\omega_\rc}
    \simeq 10 \%,
    \quad
    \begin{pmatrix*}[l]
        M = 6.2 \times 10^9 M_\odot, \\
        B = 30 ~ \si{Gauss}, \\
        g_{a\gamma} = 10^{-11} ~ \si{GeV}^{-1}    
    \end{pmatrix*}
    \label{axion_rate_0-6}
\end{align}
at the frequencies satisfying the condition for efficient conversion $\Delta_\osc /2 \simeq \Delta_\rM$, i.e., the white region in Fig.~\ref{fig:omega-ne-3}.

\subsection{Photon and axion spectra}
\label{subsec:spectra}

\subsubsection{Spherical gas model}
\label{subsubsec:spherical_gas_model}

The spectrum of axions produced by photon-axion conversion near a photon sphere can be calculated based on the formula \eqref{axion_rate_0-1}. 
Thus, the spectrum of photons incorporating dimming can be derived by subtracting the produced axion spectrum from the original photon spectrum. 

Of course, to predict a concrete shape of the spectra, we have to know the number of photons entering a region near the photon sphere for each frequency, i.e., the integrand of Eq.~\eqref{axion_rate_0-1}.
Hence, we need to identify the source of such photons.
For astrophysical black holes, the source of the photons is thought to be the surrounding gas.
In realistic situations, however, the configuration of the gas would be complicated, which makes the precise calculation of the spectra difficult.
Instead, as the simplest modeling, here we assume that the gas emitting photons is distributed in a spherical region centered at the black hole. This simple assumption allows us to calculate the spectra without relying on numerical simulations. 
Furthermore, a low radiative efficiency observed for supermassive black holes such as M87$^*$ and Sgr A$^*$ implies that their radiating region is not a simple disk but a geometrically thick hot accretion flow (see e.g.~Ref.~\cite{Yuan:2014gma}). 
Thus, we believe that the result based on our spherical gas model will give us a rough order-of-magnitude estimation for the spectra.

Under the assumption of the spherical source, the number of photons approaching a photon sphere is calculated in Eq.~\eqref{E5}:
\begin{align}
    \frac{d^3 N}{dt d\omega_\rc db}
    &= \frac{4\pi^2}{\sqrt{3}} 
    \int_{r_{\text{in}}}^{r_\text{out}} dr_\re \, J^{(N)}_\re \left( \frac{\omega_\rc}{\sqrt{3f(r_\re)}}, r_\re \right)
    \notag \\
    &\quad \times \frac{b\, r_\re}{\sqrt{ (r_\re^2 / f(r_\re)) - b^2 }} \ ,
    \label{axion_emit_1}
\end{align}
where $r_\re$ is the radial coordinate of an emission point of photons.
Here, we assumed that the emission region is a spherical shell with the inner diameter $r_\mathrm{in}$ and the outer diameter $r_{\mathrm{out}}$. 
As is defined in Eq.~\eqref{E1}, $J^{(N)}_\re$ denotes the number of emitted photons per unit time, unit frequency, unit volume, and unit solid angle.

Since we are interested in photons with impact parameter $b$ close to $b_{\crit} = 3\sqrt{3} M$, we set Eq.~\eqref{axion_emit_1} to the value at $b = b_{\crit}$ as an approximation as done in Eq.~\eqref{axion_rate_0-3}. Then, the integration with respect to $b$ over $(b_{\crit}, b_{\crit} + \sqrt{3} \epsilon^2 M /2)$, which corresponds to summing up photons entering a region $3M < r \lesssim (3+\epsilon) M$, gives
\begin{align}
    \frac{d^2 N}{dt d\omega_\rc} &\simeq 
    2 \pi^2 \epsilon^2 M 
    \int_{r_{\text{in}}}^{r_\text{out}} dr_\re \,
    J^{(N)}_\re \left( \frac{\omega_\rc}{\sqrt{3f(r_\re)}}, r_\re \right)
    \notag \\
    &\quad \times\frac{3\sqrt{3} M r_\re}{\sqrt{ (r_\re^2 / f(r_\re)) - 27 M^2}}.
    \label{original_lum}
\end{align}
Here, we assumed that $r_{\text{in}}$ is well outside the photon sphere so that $r_{\text{in}}^2 / f(r_{\text{in}}) > b^2$ holds for $b \in (b_{\crit}, b_{\crit} + \sqrt{3} \epsilon^2 M /2)$.
Multiplying Eq.~\eqref{original_lum} by an energy $\omega_\rc$, we obtain the \textit{original} photon spectral luminosity (i.e., the spectral luminosity before incorporating dimming by photon-axion conversion) in the region $3M < r \lesssim (3+\epsilon) M$.

On the other hand, by inserting Eq.~\eqref{axion_emit_1} into Eq.~\eqref{axion_rate_0-3}, we can write the number of photons converted into axions near the photon sphere per unit time and unit frequency as 
\begin{align}
    \frac{d^2N_{\gamma \to a}}{dt d\omega_\rc} 
    &\simeq \frac{\pi^2 \epsilon^2 M}{2} \left( \frac{\Delta_\rM}{\Delta_\osc / 2} \right)^2
    \frac{(3M \Delta_{\osc})^2}{1 + (3M \Delta_{\osc})^2} 
    \notag \\
    &\quad \times 
    \int_{r_{\text{in}}}^{r_{\text{out}}} dr_\re \, J_\re^{(N)} \left( \frac{\omega_\rc}{\sqrt{3 f(r_\re)}} , r_\re \right)
    \notag \\
    &\quad \times \frac{3\sqrt{3} M r_\re}{\sqrt{ (r_\re^2 / f(r_\re)) - 27 M^2}}.
    \label{axion_emit_2}
\end{align}

\subsubsection{Thermal bremsstrahlung}
\label{subsubsec:thermal_bremsstrahlung}

From Fig.~\ref{fig:omega-ne-3}, where the parameters are taken from M87$^*$, the frequency showing the efficient conversion lies in X-ray, $\omega \sim 10^{3\text{--}5} ~\si{eV}$, and gamma-ray bands, $\omega \gtrsim 10^5~\si{eV}$. For supermassive black holes, the main mechanism of such high-frequency radiation is the thermal bremsstrahlung of plasma \cite{Quataert:2002xn}.
The radiation energy by thermal bremsstrahlung per unit time, unit frequency, and unit volume is given by (see e.g.~Sec.~5.2 in Ref.~\cite{Rybicki:2004hfl})
\begin{align}
    \frac{dE}{d\tau_\re d\omega_\re dV_\re} 
    &= \frac{2^4 \alpha^3}{3 m_e} \left( \frac{2\pi}{3m_e} \right)^{1/2} T_e^{-1/2} n_e^2 e^{-\omega_\re / T_e} \bar{g}_{ff},
    \label{bre_0}
\end{align} 
where $T_e$ is the electron temperature, $n_e$ is the electron number density, and $\bar{g}_{ff}$ is a velocity averaged Gaunt factor.
Here, $\tau_\re$, $\omega_\re$, and $V_\re$ respectively denote time, frequency, and volume in a local inertial frame at the emission point.
We assumed that the ion density $n_i$ is equal to $n_e$.
Strictly speaking, $\bar{g}_{ff}$ depends on $T_e$ and $\omega_\re$, but for the order-of-magnitude estimation,  it can be regarded as unity approximately. 
Assuming isotropic radiation from each infinitesimal volume, $J_\re^{(N)}$ defined by Eq.~\eqref{E1} reads 
\begin{align}
    J_\re^{(N)}(\omega_\re, r_\re)
    = \frac{4\alpha^3}{3\pi m_e} \left( \frac{2\pi}{3m_e} \right)^{1/2} \omega_\re^{-1} T_e^{-1/2} n_e^2 e^{-\omega_\re / T_e} \bar{g}_{ff}.
    \label{bre_J}
\end{align}

Let us assume that the electron temperature and number density
obey the power law: 
\begin{align}
    T_e &= T_{e, \rc} \left( \frac{r_\re}{3M} \right)^{-p_T}, 
    \label{pow_T}
    \\
    n_e &= n_{e, \rc} \left( \frac{r_\re}{3M} \right)^{-p_n},
    \label{pow_n}
\end{align}
where $T_{e, \rc}$ and $n_{e, \rc}$ are the values at the photon sphere, and $p_T$ and $p_n$ are parameters. 
If the gas is heated to the virial temperature, $T_{\text{vir}} = m_p M / (3r) \sim 10^{12} ~ \si{K}~ (r/(3M))^{-1}$ with $m_p$ being the proton mass, we have $p_T = 1$. In theoretical models, $T_e$ is treated as a sub-virial temperature due to cooling processes and inefficient coupling between electrons and ions \cite{Yuan:2014gma}. 
In the case of spherical accretion, the mass accretion rate is written as $\dot{M} = 4 \pi r^2 \rho v_r$ with mass density $\rho$ and radial velocity $v_r$. Assuming a constant $\dot{M}$ and free falling gas $v_r \propto r^{-1/2}$, we have $p_n = 3/2$. Of course, some other factors (e.g., the presence of outflows) will modify these parameters. However,
a set of parameters $(p_T,p_n) = (1,3/2)$ is a reasonable trial.

We can perform the integral with respect to $r_\re$ in Eq.~\eqref{original_lum} with Eqs.~\eqref{bre_J}--\eqref{pow_n} in an elementary way under the following approximations. As long as the source of photons is located well outside the photon sphere, the approximation $\sqrt{ (r_\re^2 / f(r_\re)) - 27M^2 } \sim r_\re$ holds for the last line in Eq.~\eqref{original_lum}, and $f(r_\re)$ in the argument of $J_\re^{(N)}$ can be set to unity.
The exponential factor $e^{-\omega_\re / T_e}$ in Eq.~\eqref{bre_J} is also approximately unity if
\begin{align}
    \frac{\omega_\rc}{T_{e,\rc}} \ll \sqrt{3}  \left( \frac{3M}{r_\re} \right)^{p_T}.
    \label{axion_emit_3}
\end{align}
We consider $\omega_\rc$ such that this inequality is satisfied in a region outside $r_\text{in}$.
Then, Eq.~\eqref{original_lum} reduces to\footnote{The upper limit of the integration interval is approximately given by a point where the inequality \eqref{axion_emit_3} saturates, but the final result of Eq.~\eqref{axion_emit_4} depends only on the lower limit $r_\text{in}$.} 
\begin{align}
    \frac{d^2 N}{dt d\omega_\rc} &\simeq 
    27 \omega_\rc^{-1} 
    L_{\omega 0}
    \int_{r_{\text{in}}}
    \frac{dr_\re}{3M} \, \left( \frac{r_\re}{3M} \right)^{-2p_n + (p_T/2)}
    \notag \\
    &\simeq \frac{27}{2p_n - (p_T/2) - 1} \left( \frac{r_\text{in}}{3M} \right)^{-2p_n + (p_T / 2) + 1} \omega_\rc^{-1} 
    L_{\omega 0},
    \label{axion_emit_4}
\end{align}
where we defined
\begin{align}
    L_{\omega 0} 
    &\defi 2 \pi^2 \epsilon^2 M^3  
    \left( \frac{4 \alpha^3}{3 \pi m_e} \right) \left( \frac{2\pi}{3m_e} \right)^{1/2} T_{e,\rc}^{-1/2} n_{e,\rc}^2 \bar{g}_{ff}
    \notag \\
    &= 1.7 \times 10^{34} 
    \epsilon^2 \left( \frac{M}{10^9M_\odot} \right)^3
    \left( \frac{T_{e,\rc}}{10^{11} ~\si{K}} \right)^{-1/2} 
    \notag \\
    &\quad \times
    \left( \frac{n_{e,\rc}}{10^4 ~ \si{cm^{-3}}} \right)^2 
    \bar{g}_{ff}
    ~ \si{keV \cdot sec^{-1} \cdot keV^{-1}}.
    \label{axion_emit_5-2}
\end{align}
In the second line in Eq.\,\eqref{axion_emit_4}, we picked up only the term of $r_\text{in}$ by assuming $-2p_n + (p_T / 2) + 1 < 0$. 
In particular, setting $(p_T, p_n) = (1,3/2)$ and $r_\text{in} = 4M$, we have the original photon spectral luminosity near the photon sphere as $L_\omega = 11.7 L_{\omega 0}$ for $\omega_\mathrm{o} \ll  T_{e,\rc} (3M/r_\text{in}) \simeq 9 ~\si{MeV} (T_{e,\rc} /10^{11} \, \si{K})(3M / r_\text{in})$, which produces a flat spectrum at energies below gamma rays.
On the other hand, for $\omega_\mathrm{o} \gtrsim9 ~\si{MeV} (T_{e,\rc} /10^{11} \, \si{K})(3M / r_\text{in})$, the spectrum is exponentially damped since there are few high-temperature electrons which emit such high-energy photons.

\subsubsection{Observing axions through photon ring dimming}
\label{subsubsec:observing}

Inserting Eq.~\eqref{bre_J} with Eqs.~\eqref{pow_T} and \eqref{pow_n} into Eq.~\eqref{axion_emit_2}, and multiplying the energy $\omega_\rc$, we can obtain the expected spectral luminosity of axions in the model of thermal bremsstrahlung of the spherical gas. 
Thus, the spectral luminosity of photons from a region near the photon sphere can be deduced by subtracting the produced axion luminosity from the original photon luminosity.

We expect that the photon-axion conversion will affect the observed photon spectrum.
However, since the conversion occurs only near the photon sphere, we should note that only the spectrum near the photon sphere can be distorted.    
Thus, to observe the spectral distortion, we need to resolve the near-horizon region itself.
While the Event Horizon Telescope has successfully imaged the near-horizon structure in the radio band, such high-resolution observations are not currently operated in the X-ray and gamma-ray bands.
In this situation, the total luminosity from the region outside the black hole will be relevant.
When observing a black hole over a size $R(> r_{\text{in}})$, we collect the total luminosity as 
\begin{align}
    L_{\text{tot}} 
    &= 4\pi \int_{r_{\text{in}}}^{R}
    dr_\re 
    \, r_\re^2 
    \, \frac{dE}{d\tau_\re d\omega_\re dV_\re}
    \notag \\
    &\sim \frac{6^3}{\epsilon^2} \frac{1}{3-2p_n + (p_T/2)} 
    \left( \frac{R}{3M} \right)^{3-2p_n + (p_T/2)} L_{\omega 0},
    \label{totlum}
\end{align}
where we used Eq.\,\eqref{bre_0} with Eqs.\,\eqref{pow_T} and \eqref{pow_n}, and picked up the contribution around $r_\re \sim R$.
For example, by setting $p_T = 1$, $p_n = 3/2$, and $\epsilon \sim 1$, we have $L_\text{tot} \sim 4 \times 10^2 \,(R/3M)^{1/2} L_{\omega 0}$. 
This implies that the emission from the region outside the horizon will account for most of the total luminosity, so that the dimming due to the conversion will be tiny. 
Clearly, the Chandra observatory with angular resolution of arcsec is insufficient to see the dimming of M87$^*$ since it observes over a size $R \gg 3M$.

The situation will be improved if the resolution size $R$ becomes comparable to the horizon radius.
In our simple modeling, the photon sources are not distributed within $r_{\text{in}} (> 3M)$. 
Hence, if $R \lesssim 3M$, we can collect only photons approaching the photon sphere.
In this case, the total luminosity is derived from Eq.~\eqref{original_lum} [approximately Eq.\,\eqref{axion_emit_4}], and the spectral distortion becomes observable.
The required angular resolution $\theta$ is determined by $R \lesssim 3M$ as
\begin{align}
    \theta 
    = \frac{R}{D} \lesssim 10^{-5} \,\si{arcsec}\left( \frac{M}{10^{10} M_\odot} \right) \left( \frac{10\,\si{Mpc}}{D}\right),
    \label{angres}
\end{align}
where $D$ is the distance to the black hole. 
Thus, in the case of M87$^*$ ($D = 16.8\,\si{Mpc}$, $M = 6.2 \times 10^9 M_\odot$), we need the angular resolution of $\theta \lesssim 10^{-5}\,\si{arcsec}$ even in the X-ray and gamma-ray bands.

\begin{figure}
    \begin{minipage}[htb]{\hsize}
        \includegraphics[width=\hsize]{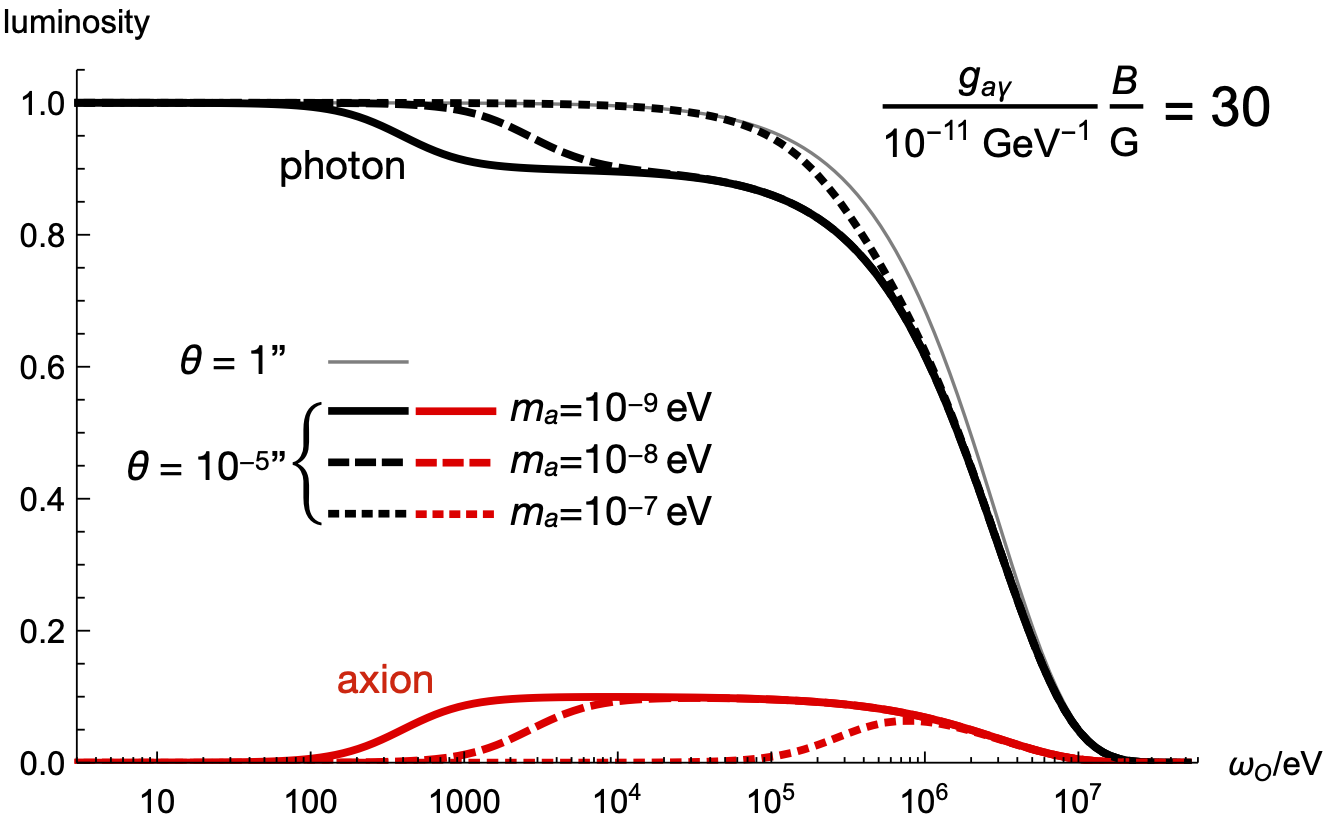}\\
        \includegraphics[width=\hsize]{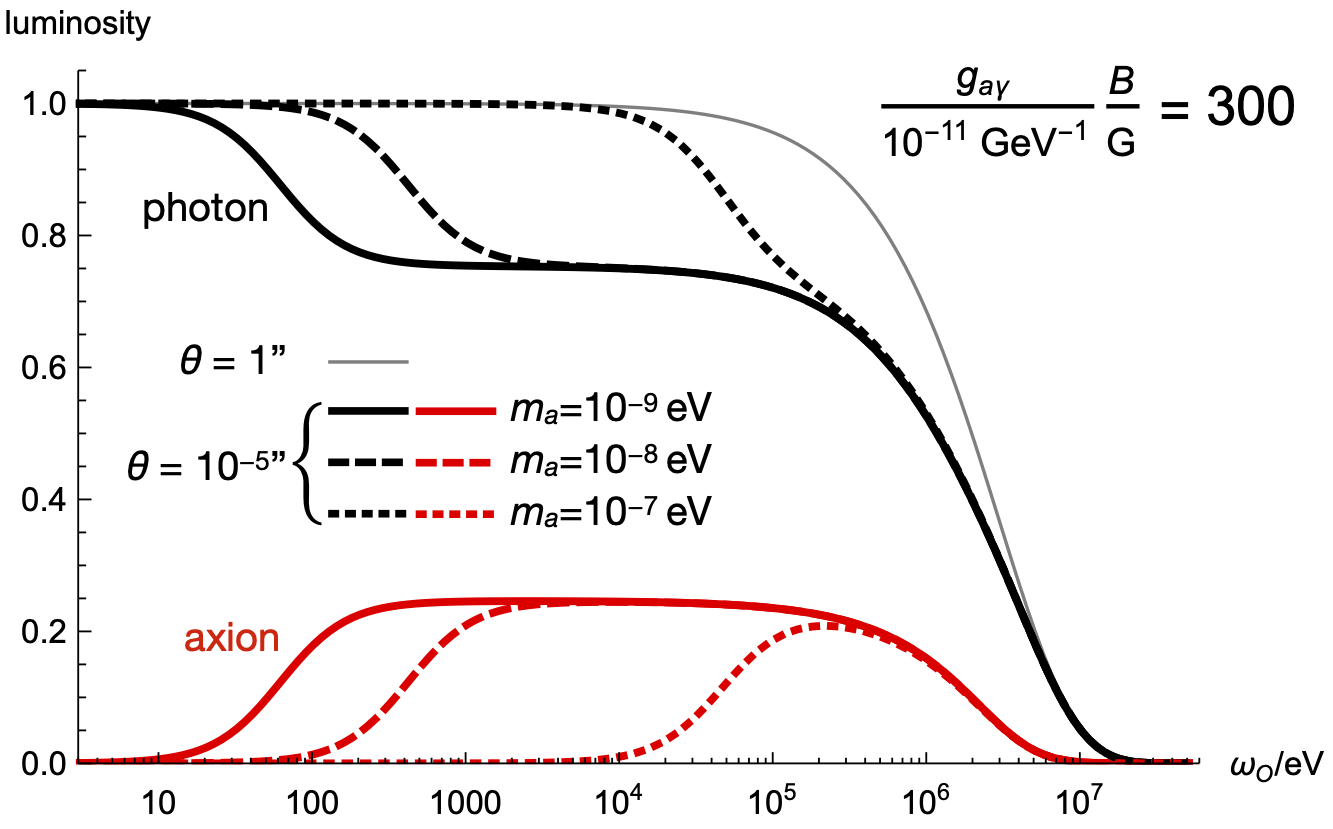}
    \end{minipage}
    \caption{The expected spectral luminosities of photons and axions are plotted.
    The horizontal axes are the observed frequency $\omega_\mathrm{o}$. 
    The thin curves are photon spectra for angular resolution of $\theta = 1\,\si{arcsec}$ like Chandra, while the black thick curves are photon spectra for angular resolution of $\theta = 10^{-5}\,\si{arcsec}$ at the Event Horizon Telescope level.
    (The target is assumed to have a distance and size of M87$^{*}$.)
    In each case, the spectral luminosity on the vertical axis is normalized by the infrared value.
    The photons are assumed to be initially produced by thermal bremsstrahlung of the gas distributed in a spherical region over $r \in (r_{\text{in}} = 4M, r_{\text{out}} = 10^3 M)$.
    The red thick curves are the axion spectral luminosities produced by the conversion from the photons, which are normalized by the infrared photon spectral luminosities at $\theta = 10^{-5}\,\si{arcsec}$.}
    For the red and black thick curves, the solid, dashed, and dotted curves are for the axion mass $m_a = 10^{-9}$, $10^{-8}$, and $10^{-7}~\si{eV}$, respectively. 
    The upper panel is for $(g_{a\gamma} / 10^{-11} \, \si{GeV}^{-1}) (B/ \si{Gauss}) = 30$, and the lower panel is for $(g_{a\gamma} / 10^{-11} \, \si{GeV}^{-1}) (B/ \si{Gauss}) = 300$.
    In both panels, other parameters are set as follows: $M=6.2 \times 10^{9}M_\odot$, $T_{e, \rc} = 10^{11}~\si{K}$, $p_T=1$, $n_{e, \rc} = 10^{4}~\si{cm^{-3}}$, and $p_n =1.5$.
    \label{fig:spectrum}
\end{figure}

Several examples of the expected energy spectra are shown in Fig.~\ref{fig:spectrum}.
There, the horizontal axes are the frequency we observe $\omega_\mathrm{o}$, which is related to that at the photon sphere $\omega_\rc$ as $\omega_\ro = \omega_\rc / \sqrt{3}$ due to gravitational redshift. We neglected other small effects such as peculiar velocities and cosmic expansion.
The thin curves are for angular resolution of $\theta = 1\,\si{arcsec}$ as in Chandra, while the thick curves are for angular resolution of $\theta = 10^{-5}\,\si{arcsec}$ at the Event Horizon Telescope level.
(The target is assumed to have a distance and size of M87$^{*}$.)
In each case, the spectral luminosity on the vertical axis is normalized by the infrared value.
We assumed that the photons are produced by thermal bremsstrahlung of the gas distributed over a spherical region $(r_{\text{in}} = 4M, r_{\text{out}} = 10^3 M)$ around the black hole. 
Originally, the photon spectrum has a cutoff around at $\omega_\ro \sim T_{e,\rc} \sim 10^7 ~\si{eV}$ due to the exponential suppression factor in Eq.~\eqref{bre_0}. 
In the case of $\theta = 1\,\si{arcsec}$, we cannot see any spectral distortion in the X-ray band due to the lack of angular resolution.
On the other hand, in the case of $\theta = 10^{-5}\,\si{arcsec}$, we resolve the photon sphere itself, so we see the dimming in the X-ray band.
Namely, the black thick curves are the photon spectra incorporating the conversion into axions around the photon sphere.
The red thick curves are the produced axion spectra.
The solid, dashed, and dotted curves are for the axion mass $m_a= 10^{-9}$, $10^{-8}$, and $10^{-7}~\si{eV}$, respectively.
In both panels in Fig.~\ref{fig:spectrum}, we set $M=6.2 \times 10^{9}M_\odot$, $T_{e, \rc} = 10^{11}~\si{K}$, $p_T=1$, $n_{e, \rc} = 10^{4}~\si{cm^{-3}}$, and $p_n =1.5$, which are taken from M87$^*$. 
The parameters in the upper panel are taken so that $(g_{a\gamma} / 10^{-11} \, \si{GeV}^{-1}) (B/ \si{Gauss}) = 30$ as in Fig.~\ref{fig:omega-ne-3}, while those in the lower panel are 
taken so that $(g_{a\gamma} / 10^{-11} \,\si{GeV}^{-1}) (B/ \si{Gauss}) = 300$ for which the conversion is more effective.

From Fig.\,\ref{fig:spectrum}, we can see that the frequency range exhibiting dimming depends on the axion mass. This is because the lowest frequency of efficient conversion is determined by the axion mass as we can see from Eq.~\eqref{efficient2}.
On the other hand, the magnitude of dimming at the efficient conversion is determined by $g_{a\gamma}$, $B$, and $M$ as shown in Eqs.~\eqref{axion_rate_0-4-2} and \eqref{axion_rate_0-5}. In the upper panel, dimming by $10\%$ can be seen, which is already mentioned in Eq.~\eqref{axion_rate_0-6}. In the lower panel, the dimming reaches around $25\%$, which is the maximum possible value.

The above demonstration shows that there is a chance to determine (or give a constraint on) the axion mass $m_a$ and axion-photon coupling $g_{a\gamma}$ by observing the photon spectrum from the vicinity of a photon sphere. Let us suppose that the mass of the black hole $M$, and the magnetic field $B$ and electron density $n_{e}$ around the photon sphere are known. The spectral shape is characterized by two quantities: the magnitude of the dimming and the frequency range exhibiting the dimming. 
First, given $M$ and $B$, the maximum magnitude of the dimming can be translated to $g_{a\gamma}$ by using Eqs.~\eqref{axion_rate_0-4-2} and \eqref{axion_rate_0-5}. Note that no matter how large $g_{a\gamma}$ is, the dimming saturates at 25\%. Hence, in the case that the dimming reaches around 25\%, only the lower bound of $g_{a\gamma}$ is determined.
Second, from Eqs.~\eqref{efficient2} and \eqref{efficient1}, the lowest frequency exhibiting the efficient dimming depends on $g_{a\gamma}$, $B$, and the axion mass $m_a$ or plasma frequency $\omega_{\Pl}$ (or, equivalently, the electron density $n_e$). In the case $m_a < \omega_\Pl$, the lowest frequency is determined by $n_e$, $g_{a\gamma}$, and $B$ as Eq.~\eqref{efficient1}, thus it is also useful to read off $g_{a\gamma}$ from the dimming. 
In a more interesting case $m_a > \omega_\Pl$, the lowest frequency is determined by Eq.~\eqref{efficient2} where $m_a$, $g_{a\gamma}$, and $B$ appear. In this case, if $B$ and $g_{a\gamma}$ are known, the lowest frequency can be used to determine $m_a$. Even if $g_{a\gamma}$ has not been determined, we can obtain the one-to-one relation between $m_a$ and $g_{a\gamma}$ through the lowest frequency exhibiting dimming. Of course, if no dimming is observed, we obtain a constraint on $m_a$ and $g_{a\gamma}$.

\subsubsection{Approximate formula for axion flux}
\label{subsubsec:formulae}

By using Eqs.~\eqref{axion_rate_0-4} and \eqref{axion_emit_4}, the spectral number flux of axions from a region near the photon sphere approximately reads 
\begin{align}
    F^{(N)}_{a, \omega}
    &= \frac{1}{4\pi D^2} \frac{d^2 N_{\gamma \to a}}{dt d\omega_\mathrm{o}}
    \notag \\
    &\simeq \frac{1}{4\pi D^2}
    \times \left[ \frac{1}{4} \left( \frac{\Delta_\rM}{\Delta_\osc / 2} \right)^2 \frac{(3M \Delta_\osc)^2}{1 + (3M\Delta_\osc)^2} \right]
    \notag \\
    &\quad\times \frac{27}{2p_n - (p_T/2) - 1} \left( \frac{r_\text{in}}{3M} \right)^{-2p_n + (p_T / 2) + 1} \omega_\mathrm{o}^{-1} 
    L_{\omega 0}
    \notag \\
    &= 1.4 \times 10^{-16} \epsilon^2 
    \left[ \frac{1}{4} \left( \frac{\Delta_\rM}{\Delta_\osc / 2} \right)^2 \frac{(3M \Delta_\osc)^2}{1 + (3M\Delta_\osc)^2} \right]
    \notag \\
    &\quad \times
    \frac{27}{2p_n - (p_T/2) - 1} \left( \frac{r_\text{in}}{3M} \right)^{-2p_n + (p_T / 2) + 1} 
    \notag \\
    &\quad \times
    \left( \frac{\si{Mpc}}{D} \right)^2 
    \left( \frac{M}{10^9M_\odot} \right)^3
    \left( \frac{T_{e,\rc}}{10^{11} ~\si{K}} \right)^{-1/2} 
    \notag \\
    &\quad \times
    \left( \frac{n_{e,\rc}}{10^4 ~ \si{cm^{-3}}} \right)^2 
    \left( \frac{\si{keV}}{\omega_\mathrm{o}} \right)
    \bar{g}_{ff}
    ~ \si{cm^{-2} \cdot sec^{-1} \cdot keV^{-1}},
    \label{axion_emit_6}
\end{align}
for $\omega_\mathrm{o} \ll 9 ~ \si{MeV} (T_{e,\rc}/ 10^{11} \, \si{K})(3M/r_\text{in})$, where $D$ is the distance to the black hole from us.
In particular, by setting the parameters following M87$^*$ ($D = 16.8 ~\si{Mpc}$) as in the upper panel in Fig.~\ref{fig:spectrum} with $m_a = 10^{-9} ~\si{eV}$, the fraction of photons converted into axions [the square bracket in Eq.~\eqref{axion_emit_6}] reaches 10\% for $\si{keV} \lesssim \omega_\mathrm{o} \lesssim 10~\si{MeV}$. Then, with $\epsilon \sim \bar{g}_{ff} \sim 1$, we have $F^{(N)}_{a, \omega} \sim 1 \times 10^{-16} \si{cm^{-2} \cdot sec^{-1} \cdot keV^{-1}}$ for $\omega_{\mathrm{o}} \sim \si{keV}$.
For the black hole at the center of the Milky Way, Sgr A$^*$ ($D = 8 ~ \si{kpc}$, $M = 4 \times 10^6 M_\odot$ \cite{EventHorizonTelescope:2022wkp}), the distance $D$ is three orders of magnitude closer than M87$^*$, but the mass $M$ is three orders of magnitude smaller. Thus, it is difficult to expect axion flux larger than M87$^*$.

\section{Conclusion}
\label{sec:conclusion}

If axions exist in nature, photons propagating in a magnetic field are converted into axions through the coupling  $(g_{a\gamma} / 4) \phi F_{\mu\nu} \wt{F}^{\mu\nu}$. As is well known, there exist black holes in the center of active galactic nuclei.  Moreover, we can expect sizable magnetic fields around black holes.
In this paper, we have investigated the photon-axion conversion phenomenon around black holes. For the magnetic field $B \sim 10^{1 \text{--} 2}~\si{Gauss}$ and axion-photon coupling $g_{a\gamma} \sim 10^{-11} ~\si{GeV}^{-1}$, the propagation length required for conversion turned out to be comparable to the Schwarzschild radius of a supermassive black hole with mass $M \sim 10^{9 \text{--} 10} M_\odot$ as shown in Eq.~\eqref{convlength}. Naively, it seems that the magnetic field has to be maintained over the conversion length in the radial direction. However, photons can orbit around a photon sphere of black holes for a certain period of time. Since such orbiting photons stay at a nearly constant radius, it is automatically ensured that the magnetic field is maintained during propagation. Thus, it is expected that photon-axion conversion will efficiently occur near the photon sphere, which will affect the observation of the photon ring around the black hole shadow.

Supposing the coupling constant $g_{a\gamma} \sim 10^{-11} ~ \si{GeV}^{-1}$, the magnetic field  $B \sim 30 ~\si{Gauss}$ and electron number density $n_e \sim 10^{4\text{--}7} ~ \si{cm^{-3}}$, which are expected values near M87$^*$ \cite{EventHorizonTelescope:2021srq}, we have shown that the photons in the X-ray and gamma-ray bands can be efficiently converted into axions in the mass range $m_a \lesssim 10^{-7}~\si{eV}$ (see Fig.~\ref{fig:omega-ne-3}). This fact indicates that, when we observe the vicinity of the black hole with electromagnetic waves, we will see a dimming of the photon ring in those wavelengths, if a sufficiently high resolution is achieved in the future.
We have shown that the maximum dimming rate of the photon ring is 25\%. In the case of M87$^*$, we found that the dimming rate could be around 10\% if the angular resolution of the Event Horizon Telescope level, i.e., $\theta \sim 10^{-5}\,\si{arcsec}$ is achieved.
In general, the magnitude of dimming depends on $g_{a\gamma}$, $M$, and $B$ as in Eqs.~\eqref{axion_rate_0-4-2} and \eqref{axion_rate_0-5}. We depicted the dependence in Fig.~\ref{fig:dimming}. Provided $M$ and $B$ are known from other observations, we can determine the value of (or a constraint on) $g_{a\gamma}$ from photon ring dimming. The larger $M$, the greater the dimming, thus supermassive black holes such as M87$^*$ are good candidates for observing photon ring dimming.
Furthermore, the frequency range exhibiting the dimming has information on the axion mass. In the case that the axion mass $m_a$ is smaller than the plasma frequency $\omega_{\Pl}$, the lowest frequency of dimming is determined by Eq.~\eqref{efficient1} where $g_{a\gamma}$, $n_e$, and $B$ appear. On the other hand, in the case $m_a > \omega_{\Pl}$, it is determined by Eq.~\eqref{efficient2} where 
$g_{a\gamma}$, $m_a$, and $B$ appear. Hence, the lowest frequency of dimming can be used to measure $g_{a\gamma}$ and $m_a$. We have demonstrated photon ring dimming in Fig.~\ref{fig:spectrum}, where the photons are assumed to be sourced by thermal bremsstrahlung of gas spherically distributed around the black hole.
Of course, to see the photon ring dimming, it is necessary to observe the near-horizon region with much high resolution. While the Event Horizon Telescope has succeeded in imaging the region in the radio band, such high-resolution observations have not yet been achieved in the X-ray and gamma-ray bands.
The present study anticipates the future potential of multi-wavelength observations with higher resolution.
As to this direction, we refer the reader to Ref.~\cite{Uttley:2019ngm} which proposed  high-resolution X-ray interferometry.
We believe that our results will give  motivation for future observations.

In this paper, the magnetic field and the plasma density are treated as homogeneous near the photon sphere for simplicity. 
Interestingly, from Fig.~\ref{fig:omega-ne-3}, we can see that conversion occurs in the large parameter region. Therefore, we expect that conversion will occur even in more realistic cases, namely, inhomogeneous magnetic fields and inhomogeneous plasma density.

There are several directions to be pursued beyond the present work. One is to include the rotation of the black hole. Interestingly, the Kerr black hole has circular orbits of photons at two different radii on the equatorial plane. Since the photons emitted from the different radii undergo different gravitational redshifts, conversion into axions at those radii may create dimming at different frequencies. Another important issue is to study the effect of photon-axion conversion on the polarization of light coming from the photon sphere.
It is also worth studying conversion not only in the background of the magnetic field but also in the background of the axion~\cite{Masaki:2019ggg}, while only the former is considered in this paper. In fact, axions could be dark matter \cite{Preskill:1982cy, Marsh:2015xka, Hui:2016ltb, Chadha-Day:2021szb} or produced by superradiance around black holes \cite{Arvanitaki:2009fg, Brito:2015oca}.
Finally, we have estimated the axion flux from the photon sphere of a single black hole in Eq.~\eqref{axion_emit_6}, which is too tiny to be detected. However, there are a huge number of black holes in our universe. Especially, quasars are significantly brighter than low-luminosity active galactic nuclei such as M87$^*$, and resulting axion luminosity may also be larger than that of M87$^*$. The sum of these contributions will make up a part of the cosmic axion background \cite{Dror:2021nyr}. It will be intriguing to evaluate these contributions.
We leave these issues for future work.

\begin{acknowledgments}
K.N.~thanks Chul-Moon Yoo and Che-Yu Chen for useful comments.
K.N.~was supported by Grant-in-Aid for JSPS Research Fellowship and JSPS KAKENHI Grant No.~JP21J20600.
K.S.~was supported by JST SPRING, Grant Number JPMJSP2148.
J.S.~was in part supported by JSPS KAKENHI Grant Numbers JP17H02894, JP17K18778, JP20H01902, JP22H01220.
\end{acknowledgments}

\appendix

\section{Derivation of the photon-axion conversion probability}
\label{app:conversion}

In this Appendix, we study the conversion phenomenon between photons and axions in a constant external magnetic field. Our purpose is to derive a formula for the conversion probability
in flat space. (The derivation has been done e.g.~in Refs.~\cite{Raffelt:1987im, Hochmuth:2007hk, Masaki:2017aea}.)
Throughout this Appendix, we set $\hbar = c = 1$, and use rationalized Heaviside--Lorentz units for electromagnetism, where $4\pi$ does not appear in the Maxwell equations but does in the Coulomb law. 

We consider a system of the axion and photon, 
\begin{align}
    \mathcal{L} &= - \frac{1}{4} F_{\mu\nu} F^{\mu\nu} - \frac{1}{2} \pd_\mu \phi \pd^\mu \phi 
    - \frac{1}{2} m_a^2 \phi^2 - \frac{1}{4} g_{a\gamma} \phi F_{\mu\nu} \wt{F}^{\mu\nu} ,
\end{align}
where $\phi$ is the axion field with mass $m_a$, $g_{a\gamma}$ is the axion-photon coupling constant, $F_{\mu\nu}$ is the electromagnetic field strength tensor, and $\wt{F}_{\mu\nu}$ is the dual of $F_{\mu\nu}$ given by $\wt{F}_{\mu\nu} = (1/2) \epsilon_{\mu\nu\rho\sigma} F^{\rho\sigma}$ with $\epsilon_{\mu\nu\rho\sigma}$ being completely anti-symmetric in its indices and normalized as $\epsilon_{0123} = 1$.

The equations of motion for the axion and photon are 
\begin{align}
    &\square \phi - m_a^2 \phi = \frac{1}{4} g_{a\gamma} F_{\mu\nu} \wt{F}^{\mu\nu} ,
    \label{eomaxi}
\end{align}
and 
\begin{align}
    \pd_{\mu} F^{\mu\nu} = - g_{a\gamma} \wt{F}^{\mu\nu}  \pd_\mu \phi,
    \label{eomA1}
\end{align}
respectively,\footnote{In the derivation, the Bianchi identity $\pd_\mu \wt{F}^{\mu\nu} = 0$ is used.} where we defined $\square = \pd^\mu \pd_\mu$.

We consider a situation where electromagnetic waves propagate in the background of a constant magnetic field.
The electromagnetic field is the sum of the magnetic field and electromagnetic waves:
\begin{align}
    F_{\mu\nu} = \bar{F}_{\mu\nu} + \pd_\mu A_\nu - \pd_\nu A_\mu.
    \label{mag}
\end{align}
The background magnetic field $\bar{F}_{\mu\nu}$
is represented by
\begin{align}
    \bar{F}_{0i} &=0, \\
    \wt{\bar{F}}_{0i} &= \frac{1}{2} \epsilon_{0ijk} \bar{F}_{jk} = B_i (=\text{constant}).
\end{align}
For the propagating photons $A_\mu$, we choose the Coulomb gauge condition,
\begin{align}
    \bm{\nabla} \cdot \textbf{A} = 0.
\end{align}
Hereafter, we use equations in the linear order of $A_\mu$ or $\phi$.
Under the Coulomb gauge, the spatial components of Eq.~\eqref{eomA1} read
\begin{align}
    \square \mathbf{A} - \bm{\nabla} \dot{A}^0 = g_{a\gamma} \mathbf{B} \dot{\phi},
    \label{eomA2-2}
\end{align}
where a dot represents a time derivative.
The $A^0$ component is determined by a constraint equation following from the $\nu = 0$ component of Eq.~\eqref{eomA1},
\begin{align}
    \bm{\nabla}^2 A^0 = - g_{a\gamma} \mathbf{B} \cdot \bm{\nabla} \phi.
    \label{eomA2-3}
\end{align}
On the other hand, the equation of motion of the axion \eqref{eomaxi} is recast to
\begin{align}
    (\square - m_a^2) \phi = - g_{a\gamma} \mathbf{B} \cdot (\dot{\mathbf{A}} + \bm{\nabla} A^0).
    \label{eomaxi2}
\end{align}
From Eqs.~\eqref{eomA2-2} and \eqref{eomaxi2}, it can be seen that only the component of $\mathbf{A}$ parallel to $\mathbf{B}$ has mixing with the axion. For simplicity, let us take $\mathbf{A}$ and $\phi$ to be plane waves propagating along the $z$-direction. The $A_z$ component vanishes because of the Coulomb gauge condition. Without loss of generality, we can take $\mathbf{B}$ to lie in the $x$-$z$ plane. Thus, in the $(x,y,z)$ coordinates, we set\footnote{Note that $B\sin \Theta$ here is simply denoted by $B$ in the main sections.}
\begin{align}
    \mathbf{B} &= (B \sin \Theta, 0, B \cos \Theta), \\
    \mathbf{A} &= (iA_\parallel (t,z), iA_\perp(t,z), 0),
    \label{Acomp}
\end{align}
where $\Theta$ is the angle between the direction of $\mathbf{B}$ and the $z$-axis (the direction of the wave number vector), and the factor $i$ in the definition of $A_\parallel$ and $A_\perp$ is put for later convenience. 
In the leading (free field) approximation, $A_\parallel$ and $A_\perp$ have plane wave solutions $\propto e^{-i(\omega t - kz)}$ with $k = \omega$, 
and we can safely set $A^0 = 0$.
Furthermore, when we consider relativistic axions with momentum $k \gg m_a$, it is also a good approximation to take the axions as the plane wave $\propto e^{-i(\omega t - kz)}$ with $k = \omega$ at the leading order.

In the presence of the magnetic field, we need to consider the Euler--Heisenberg effective Lagrangian, 
\begin{align}
    \mathcal{L}_{\text{EH}} = \frac{\alpha^2}{90m_e^4} \left( (F_{\mu\nu} F^{\mu\nu})^2 + \frac{7}{4} (F_{\mu\nu} \wt{F}^{\mu\nu})^2 \right),
\end{align}
where $\alpha = 1/137$ is the fine-structure constant and $m_e = 511 ~\si{keV}$ is the electron mass. This Lagrangian induces the following term
\begin{align}
    \frac{4 \alpha^2}{45 m_e^4} \pd_\mu \left( F_{\rho\sigma} F^{\rho\sigma} F^{\mu\nu} + \frac{7}{4} F_{\rho\sigma} \wt{F}^{\rho\sigma} \wt{F}^{\mu\nu} \right)
    \label{EH2}
\end{align}
into the right-hand side of Eq.~\eqref{eomA1}.
Using the parametrization \eqref{Acomp} and assuming the plane wave solution, we have 
\begin{align}
    F_{\mu\nu} F^{\mu\nu} &= 2B^2 + 4 \omega B \sin \Theta A_\perp, \\
    F_{\mu\nu} \wt{F}^{\mu\nu} &= - 4 \omega B \sin \Theta A_\parallel,
\end{align}
in the linear order of $\mathbf{A}$.
Then, at the leading approximation, the $\nu = x$ component of Eq.~\eqref{EH2} reads
\begin{align}
    -i \omega^2 \frac{4\alpha^2}{45 m_e^4} 7 (B\sin\Theta)^2 A_\parallel, 
\end{align}
and the $\nu = y$ component reads 
\begin{align}
    -i \omega^2 \frac{4\alpha^2}{45 m_e^4} 4 (B \sin \Theta)^2 A_\perp.
\end{align}
By adding these terms to the right-hand side of Eq.~\eqref{eomA2-2}, we have  
\begin{align}
    \square A_\parallel + 7 \omega^2 \xi \sin^2 \Theta A_\parallel + \omega g_{a\gamma} B \sin \Theta \phi &= 0, 
    \label{eomA5}\\
    \square A_\perp + 4 \omega^2 \xi \sin^2 \Theta A_\perp &= 0,
    \label{eomA6}
\end{align} 
where 
\begin{align}
    \xi = \frac{\alpha}{45 \pi} \left( \frac{B}{B_{\text{crit}}} \right)^2 , \quad B_\text{crit} \defi \frac{m_e^2}{e},
\end{align}
and now $\square = -\pd_t^2 + \pd_z^2$.
Hereafter, we concentrate only on $A_\parallel$ to see mixing with the axion.
The effects of surrounding plasma can be incorporated by adding a term $-\omega_\Pl^2 A_\parallel$ to the equation, where $\omega_\Pl$ denotes the plasma frequency:
\begin{align}
    \square A_\parallel - \omega_\Pl^2 A_\parallel + 7 \omega^2 \xi \sin^2 \Theta A_\parallel + \omega g_{a\gamma} B \sin \Theta \phi &= 0.
\end{align}
The equation of motion of the axion \eqref{eomaxi2} is now given by 
\begin{align}
    (\square - m_a^2) \phi + \omega g_{a\gamma} B \sin \Theta A_\parallel = 0.
    \label{eomaxi3}
\end{align}

To see the conversion, it is convenient to express $A_\parallel$ and $\phi$ as 
\begin{align}
    A_\parallel(t,z) &= \wt{A}(z) e^{-i(\omega t - kz)} + \text{h.c.}, \\
    \phi(t,z) &= \wt{\phi}(z) e^{-i(\omega t - kz)} + \text{h.c.}, \quad 
    \omega = k.
\end{align}
The plane wave $e^{-i(\omega t - kz)}$ solves the free and massless wave equations, $\square A_\parallel (t,z) = \square \phi(t,z) = 0$.
We investigate how the photons propagating over a distance in the $z$-direction convert into axions.
We have taken into account the $z$-dependent amplitudes, $\wt{A}(z)$ and $\wt{\phi}(z)$, in order to see how these amplitudes vary depending on the distance $z$.
It is expected that the variation of these amplitudes is slow in the sense that $|\pd_z^2 \wt{A}(z)| \ll k |\pd_z \wt{A}(z)| $ and $|\pd_z^2 \wt{\phi}(z)| \ll k |\pd_z \wt{\phi}(z)| $.
Thus, we have
\begin{align}
    \square A_\parallel(t,z) &\simeq 2i \omega \pd_z \wt{A}(z) e^{-i(\omega t - kz)} + \text{h.c.}, \\
    \square \phi(t,z) &\simeq 2i \omega \pd_z \wt{\phi}(z) e^{-i(\omega t - kz)} + \text{h.c.}
\end{align}
These are the lowest order approximations to see the $z$-dependence of $\wt{A}$ and $\wt{\phi}$.
Now, the equations of motion reduce to 
\begin{align}
    i \frac{d}{dz} 
    \begin{pmatrix}
        \wt{A}(z) \\
        \wt{\phi}(z)    
    \end{pmatrix}
    &= \begin{pmatrix}
        \Delta_{\Pl} - \Delta_{\vac} & - \Delta_\rM \\
        - \Delta_\rM &\Delta_a
    \end{pmatrix}
    \begin{pmatrix}
        \wt{A}(z) \\
        \wt{\phi}(z)    
    \end{pmatrix},
\end{align}
where $\Delta_\rM$, $\Delta_a$, $\Delta_{\Pl}$, and $\Delta_{\vac}$ are defined as follows:
\begin{align}
    \Delta_\rM &= \frac{1}{2} g_{a \gamma} B \sin \Theta ,
    \label{Delta_M} \\
    \Delta_a &= \frac{m_a^2}{2 \omega},
    \label{Delta_a} \\
    \Delta_{\Pl} &= \frac{\omega_{\Pl}^2}{2\omega},
    \label{Delta_pl} \\
    \Delta_{\vac} &= \frac{7}{2} \omega \frac{4\alpha^2}{45m_e^4} (B \sin \Theta)^2 .
    \label{Delta_vac} 
\end{align}
It is convenient to rewrite the equation as
\begin{align}
    i \frac{d}{dz} \vec{\Psi}(z) &= \mathbf{M} \vec{\Psi}(z) ,
    \label{diffeq}
\end{align}
where we used the notations
\begin{align}
   & \vec{\Psi}(z) = \begin{pmatrix}
        \wt{A}(z) \\
        \wt{\phi}(z)    
    \end{pmatrix}, \\
   & \mathbf{M} = \begin{pmatrix}
        \Delta_{\parallel} & - \Delta_\rM \\
        - \Delta_\rM &\Delta_a
    \end{pmatrix},
    \quad 
    \Delta_{\parallel} = \Delta_{\Pl} - \Delta_{\vac}.
\end{align}
The eigenvalues of the matrix $\mathbf{M}$ are 
\begin{align}
    \lambda_\pm = \frac{\Delta_\parallel + \Delta_a \pm \sqrt{(\Delta_\parallel - \Delta_a)^2 + 4 \Delta_\rM^2}}{2}.
\end{align}
Since $\mathbf{M}$ is a real and symmetric matrix, it can be diagonalized by an orthogonal matrix $\mathbf{R}$:
\begin{align}
    \mathbf{R}^\trans \mathbf{M} \mathbf{R} = \begin{pmatrix}
        \lambda_+ &0\\
        0 &\lambda_-
    \end{pmatrix},
    \quad 
    \mathbf{R} = \begin{pmatrix}
        \cos \theta &-\sin \theta \\
        \sin \theta &\cos \theta
    \end{pmatrix}.
\end{align}
The direct calculation gives the off-diagonal component of $\mathbf{R}^\trans \mathbf{M} \mathbf{R}$ as
\begin{align}
    &\mathbf{R}^\trans \mathbf{M} \mathbf{R}
    \notag \\
    &= \begin{pmatrix}
        * &* \\
        \cos 2\theta (-\Delta_\rM) + \frac{1}{2} \sin 2\theta (\Delta_a - \Delta_\parallel) &*
    \end{pmatrix},
\end{align}
which must vanish. Thus, the \textit{mixing angle} $\theta$ is determined as 
\begin{align}
    \tan 2\theta = \frac{2 \Delta_\rM}{\Delta_a - \Delta_\parallel}.
    \label{angle}
\end{align}
Using the matrix $\mathbf{R}$, Eq.~\eqref{diffeq} reduces to 
\begin{align}
    i \frac{d}{dz} (\mathbf{R}^\trans \vec{\Psi}(z)) = \begin{pmatrix}
        \lambda_+ & 0\\
        0& \lambda_-
    \end{pmatrix}
    (\mathbf{R}^\trans \vec{\Psi}(z)).
\end{align}
It is easy to solve this equation as 
\begin{align}
    \vec{\Psi}(z)
    = \mathbf{R} \begin{pmatrix}
        e^{-i\lambda_+ z} &0\\
        0& e^{-i\lambda_- z} 
    \end{pmatrix}
    \mathbf{R}^\trans \vec{\Psi}(0).
\end{align}
Finally, we obtain the general solutions 
\begin{align}
    \wt{A}(z)
    &= (\cos^2 \theta e^{-i\lambda_+ z} + \sin^2 \theta e^{-i\lambda_- z} ) \wt{A}(0)
    \notag \\
    &\quad + \sin \theta \cos \theta (e^{-i\lambda_+ z } - e^{-i\lambda_- z} ) \wt{\phi}(0),
    \\
    \wt{\phi}(z)
    &= \sin \theta \cos \theta (e^{-i\lambda_+ z} - e^{-i\lambda_- z}) \wt{A}(0) 
    \notag \\
    &\quad + (\sin^2 \theta e^{-i\lambda_+ z} + \cos^2\theta e^{-i\lambda_- z}) \wt{\phi}(0).
\end{align}
Given $\wt{\phi}(0) = 0$ and $\wt{A}(0) = 1$ as the initial condition, we can obtain the conversion probability at a distance $z$ as 
\begin{align}
    P_{\gamma \to a}(z)
    &= |\wt{\phi}(z)|^2 
    \notag \\
    &= \sin^2 2\theta \sin^2 \left( \frac{\Delta_\osc}{2} z \right)
    \notag \\
    &= \left( \frac{\Delta_\rM}{\Delta_\osc / 2} \right)^2 \sin^2 \left( \frac{\Delta_\osc}{2} z \right),
\end{align}
where we used Eq.~\eqref{angle} and defined 
\begin{align}
    \Delta_\osc^2 = (\Delta_\Pl - \Delta_\vac - \Delta_a)^2 + 4 \Delta_\rM^2 \ .
\end{align}

\section{Orbiting time of a photon around a photon sphere}
\label{app:orbiting_time}

In this Appendix, we consider a geodesic of a photon in Schwarzschild spacetime, and give a formula for the orbiting time of a photon around a photon sphere in terms of the impact parameter of the photon incident on the black hole.
The discussion here is based on Sec.~VII in Ref.~\cite{Yoshino:2019qsh}.
The Schwarzschild metric is given by 
\begin{align}
    ds^2 &= - f(r) dt^2 + \frac{1}{f(r)} dr^2 + r^2 (d\theta^2 + \sin^2 \theta d\phi^2), 
    \notag \\
    f(r) &= 1 - \frac{2M}{r}.
    \label{B1-2}
\end{align}
Let us consider a geodesic of a photon denoted by $x^\mu(\lambda) = (t(\lambda), r(\lambda), \theta(\lambda), \phi(\lambda))$ in the Schwarzschild coordinates, where $\lambda$ is an affine parameter. Since the Schwarzschild spacetime has spherical symmetry, we can take the geodesic on an equatorial plane $\theta = \pi/2$ without loss of generality. 
The timelike and rotational Killing vectors $\pd_t$ and $\pd_\phi$ lead to two conserved quantities along the geodesic, 
\begin{align}
    E &\defi - (\pd_t)_\mu \frac{dx^\mu}{d\lambda} = f(r) \frac{dt}{d\lambda},
    \label{B1}
    \\
    L &\defi (\pd_\phi) _\mu \frac{dx^\mu}{d\lambda} = r^2 \frac{d\phi}{d\lambda}.
    \label{B2}
\end{align}
The conserved quantities $E$ and $L$ are the energy and the angular momentum of the photon, respectively. The impact parameter of the incident photon to the black hole is defined by 
\begin{align}
    b \defi \frac{L}{E}.
    \label{B3}
\end{align}
Using the null condition of the geodesic, $g_{\mu\nu} (dx^\mu / d\lambda) (dx^\nu / d\lambda) = 0$, together with Eqs.~\eqref{B1} and \eqref{B2}, we have 
\begin{align}
    \frac{dr}{dt} &= \pm f(r) \sqrt{ 1 - \frac{b^2f(r)}{r^2} }, 
    \label{B4}\\
    \frac{d\phi}{dt} &= \frac{bf(r)}{r^2}.
    \label{B5}
\end{align}
In Eq.~\eqref{B4}, the signs $+$ and $-$ represent the outward and inward photons, respectively.

In a particular case $b = b_\crit \defi 3\sqrt{3} M$, Eq.~\eqref{B4} can be integrated analytically as
\begin{align}
    t = \pm F(r)
    \label{B6}
\end{align}
where we defined the function
\begin{align}
    F(r) 
    &= -3\sqrt{3} M \ln \bigg| \frac{\sqrt{3r} + \sqrt{r+6M}}{\sqrt{3r} - \sqrt{r+6M}} \bigg| + \sqrt{r(r+6M)} 
    \notag \\
    &\quad + 4M \ln \left( \sqrt{\frac{r}{M}} + \sqrt{\frac{r}{M} + 6}  \right) 
    \notag \\
    &\quad + 2M \ln \left( \frac{2\sqrt{r} + \sqrt{r+6M}}{2\sqrt{r} - \sqrt{r+6M}} \right).
    \label{B8}
\end{align}
From the first term of the right-hand side of Eq.~\eqref{B8}, we can see that the photon can travel for an infinitely long time on $r = 3M$, which corresponds to the photon sphere.
For later convenience, we define a function $\wt{F}(t)$ by inverting the equation $t = F(r)$ for $r > 3M$ as 
\begin{align}
    r - 3M = \wt{F}(t),
    \label{B8-2}
\end{align}
which describes the outward photon with $b = b_\crit$ outside the photon sphere.
In the neighborhood of $r=3M$, approximately we have
\begin{align}
    \wt{F}(t) \simeq M
    \exp \left( \frac{t - C}{3\sqrt{3} M} \right),
    \label{B9}
\end{align}
where a constant $C$ is given by 
\begin{align}
    C &= 3 \sqrt{3} M + 4M \ln (3 + \sqrt{3}) + 2M \ln \left( \frac{2 \sqrt{3} + 3}{2\sqrt{3} - 3} \right) 
    \notag \\
    &\quad - 3\sqrt{3} M \ln 18.
    \label{B10}
\end{align}
On the other hand, the trajectory of the inward photon with $b = b_\crit$ outside the photon sphere can be obtained by flipping the sign of $t$ in Eq.~\eqref{B8-2}.

Now let us consider a photon with impact parameter $b$ slightly larger than $b_\crit$ coming from outside the black hole. We can expect that such a photon will be approaching the photon sphere, then orbiting near the sphere, and finally escaping from the sphere. 
We model such a trajectory by the formula 
\begin{align}
    r - 3M = \wt{F}(-t - D) + \wt{F}(t).
    \label{B12}
\end{align}
In the early time, $\wt{F}(-t-D)$ is dominant so that it represents the incident inward photon. On the other hand, in the late time, $\wt{F}(t)$ is dominant so that it represents the photon escaping away from the black hole. The constant $D$ determined by the impact parameter $b$ characterizes the time 
during which the photon is staying near the photon sphere. 

To determine $D$, first note that Eq.~\eqref{B12} is approximately given by
\begin{align}
    r - 3M \simeq M \exp\left( \frac{-t-C-D}{3\sqrt{3}M} \right) + M\exp\left( \frac{t-C}{3\sqrt{3}M} \right) 
    \label{B13}
\end{align}
in the neighborhood of $r=3M$.
The minimum of the right-hand side determines the pericenter of the photon's trajectory, $r_\mathrm{p}$, as 
\begin{align}
    r_\mathrm{p} - 3M 
    = 2 M \exp \left( \frac{-C -(D/2)}{3 \sqrt{3} M} \right).
    \label{B14}
\end{align}
On the other hand, the pericenter $r_\mathrm{p}$ can be given in terms of $b$ by finding the zero of $dr/dt$ given by Eq.~\eqref{B4}.
Introducing $\delta r_\mathrm{p} = r_\mathrm{p} - 3M$ and $\delta b = b- b_\crit (>0)$, we can perturbatively solve the equation $dr/dt = 0$ as 
\begin{align}
    \frac{\delta r_\mathrm{p}}{3M} = \sqrt{ \frac{2}{3} \frac{\delta b}{b_{\crit}}} + \mathcal{O} \left( \frac{\delta b}{b_{\crit}} \right).
    \label{B15}
\end{align}
Inserting Eq.~\eqref{B15} into Eq.~\eqref{B14}, we can write $D$ in terms of $b$ as 
\begin{align}
    D = -2C - 3\sqrt{3} M \ln \frac{b-b_\crit}{\sqrt{12}M} ,
    \label{B16}
\end{align}
where we neglected the quantities with the order of $M \mathcal{O}(\sqrt{\delta b / b_{\crit}})$.

The first term of the right-hand side in Eq.~\eqref{B13} becomes less than $\epsilon M/2$ (where $\epsilon > 0$) for $t > -C-D - 3\sqrt{3} M \ln (\epsilon / 2)$, while the second term becomes less than $\epsilon M/2$ for $t < C + 3\sqrt{3} M \ln (\epsilon / 2)$. Thus, we can expect that the photon remains in a region $3M < r < (3 + \epsilon) M$ for the time interval 
\begin{align}
    T(b) &= 2C + D + 6 \sqrt{3} M \ln \frac{\epsilon}{2}
    \notag \\
    &= - 3\sqrt{3} M \ln \frac{2(b-b_\crit)}{\sqrt{3}\, \epsilon^2 M}.
    \label{B17}
\end{align}

Given Eq.~\eqref{B17}, we can find an impact parameter at which the time interval within $3M < r < (3 + \epsilon) M$ vanishes as $b = b_{\crit} + \epsilon^2 \sqrt{3} M / 2$. On the other hand, from Eq.~\eqref{B4}, the impact parameter with pericenter $r = (3+\epsilon)M$ turns out to be $b = \sqrt{(3+\epsilon)^3 / (1+ \epsilon)} M = b_\crit + \epsilon^2 \sqrt{3} M / 2 - \epsilon^3 4 M / (3\sqrt{3}) + \mathcal{O}(\epsilon^4)$, and photons with larger $b$ than this value cannot enter the region  $r < (3 + \epsilon) M$. This difference is originated from the error of $T(b)$ of $M \mathcal{O}(\sqrt{\delta b / b_{\crit}})$.
However, for a small $\epsilon$, this difference becomes negligible so that we can regard Eq.~\eqref{B17} as a good approximation for the orbiting time interval in a region $3M < r \lesssim (3+\epsilon) M$.
In fact, the formula \eqref{B17} reproduces the result obtained by solving the geodesic equation numerically \cite{Yoshino:2019qsh}.

\section{Emission angle and impact parameter}
\label{app:emission_angle}

\begin{figure}
    \includegraphics[width=\hsize]{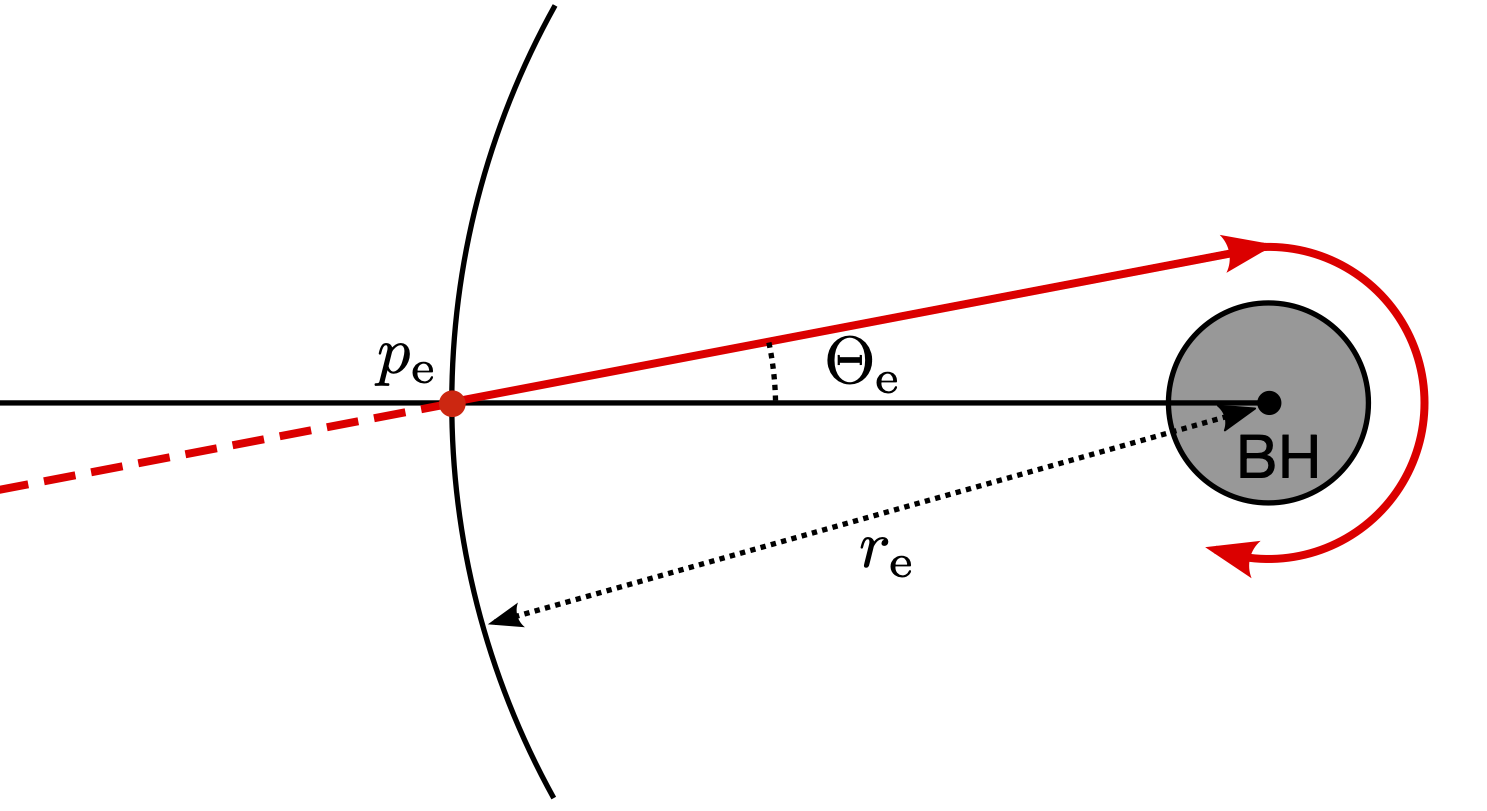}
    \caption{A light ray emitted from a point $p_\re$ toward a photon sphere of a black hole is schematically shown by a red curve. The point $p_\re$ is located on a sphere with the radius $r_\re$ centered at the black hole in the Schwarzschild coordinates. On the point $p_\re$, the light ray is emitted with the zenith angle $\Theta_\re$ measured from the line connecting $p_\re$ and the center of the black hole.}
    \label{fig:photon_angle}
\end{figure}

Let us consider a light ray emitted from a point $p_\re$, which is located at $r = r_\re$ in the Schwarzschild coordinates, toward the black hole photon sphere with impact parameter $b$ defined by Eq.~\eqref{B3}. To describe the trajectory of the photon, we introduce an angle $\Theta_\re$ between the initial direction of the incident photon and the direction to the center of the black hole in the local inertial frame at $p_\re$ as shown in Fig.~\ref{fig:photon_angle}.
More specifically, we introduce a tetrad 
\begin{align}
    e_0 &= \frac{1}{\sqrt{f(r)}} \pd_t, \\
    e_1 &= \sqrt{f(r)} \pd_r, \\
    e_2 &= \frac{1}{r} \pd_\theta, \\
    e_3 &= \frac{1}{r \sin \theta} \pd_\phi,
\end{align}
and its dual 
\begin{align}
    e^0 &= \sqrt{f(r)} dt, 
    \label{C5}\\
    e^1 &= \frac{1}{\sqrt{f(r)}} dr, \\
    e^2 &= r d\theta, \\
    e^3 &= r \sin \theta d\phi,
\end{align}
where $f(r)$ is defined in Eq.~\eqref{B1-2}.
For a while, we take the trajectory to lie on the $\theta = \pi/2$ plane in the Schwarzschild coordinates.
Then, $e_1$ and $e_3$ are orthonormal bases parallel and normal to the direction to the center of the black hole, respectively. Thus, the angle $\Theta_\re$ is given by 
\begin{align}
    \tan \Theta_\re &= \left| \frac{k^\mu (e^3)_\mu}{k^\nu (e^1)_\nu} \right|_{p_\re}
    \notag \\
    &= \left| r \sqrt{f(r)}\frac{d\phi}{dr} \right|_{p_\re},
\end{align}
where $k^\mu = dx^\mu(\lambda) / d\lambda$ is the tangent vector of the geodesic with the affine parameter $\lambda$.
By virtue of Eqs.~\eqref{B4} and \eqref{B5}, we have 
\begin{align}
    b = \frac{r_\re}{\sqrt{f(r_\re)}} \sin \Theta_\re.
    \label{C10}
\end{align}
This formula relates the emission angle $\Theta_\re$ and radial coordinate of the emission point $r_\re$ with the impact parameter $b$.

\section{Flow into a photon sphere from a spherical region}
\label{app:photon_flow_spherical}

Let us imagine a spherical region centered at a black hole, in which photons are emitted isotropically from each point with a certain emission rate.
In this Appendix, we will estimate how many of such emitted photons can approach a photon sphere of the black hole.
For simplicity, here we model the geometry to be Schwarzschild spacetime neglecting the rotation of the black hole, for which the metric is given by Eq.~\eqref{sch_metric}.

We begin by considering an infinitesimal volume $dV_\re$ at a point $p_\re$, which is located at the radial (Schwarzschild) coordinate $r_\re$.
Let us write the number of photons within a frequency width $d\omega_\re$ emitted from $dV_\re$ and passing through an infinitesimal solid angle $d\Omega_\re$ viewed from the emission point $p_\re$ per unit time $\tau_\re$ as\footnote{``$d^6$'' in the left-hand side stands for the dimension of the infinitesimal volume of the right-hand side. Here, we treat $d\Omega_\re$ and $dV_\re$ as two- and three-dimensional infinitesimal volume elements, respectively.} 
\begin{align}
    d^6 \left(\frac{dN}{d\tau_\re} \right) = J^{(N)}_\re(\omega_\re, r_\re) d\Omega_\re dV_\re d\omega_\re.
    \label{E1}
\end{align}
Here, the frequency $\omega_\re$, time $\tau_\re$, solid angle $\Omega_\re$, and volume $V_\re$ are measured in a local inertial frame at $p_\re$. In that frame with the origin at $p_\re$, we take $\Theta_\re$ to denote the zenith angle measured from the direction to the black hole (see Fig.~\ref{fig:photon_angle}), and $\Phi_\re$ to denote the azimuth angle in the plane normal to the direction to the black hole.
We assume that the emission from a point $p_\re$ is isotropic so that $J_\re^{(N)}$ does not depend on $\Theta_\re$ and $\Phi_\re$.

It is convenient to rewrite Eq.~\eqref{E1} in terms of the impact parameter of a photon $b$.
For this purpose, we can use Eq.~\eqref{C10}, and $db = (r_\re / \sqrt{f(r_\re)}) \cos \Theta_\re d\Theta_\re$ which follows from Eq.~\eqref{C10} for a fixed $r_\re$. 
Using these relations, and integrating Eq.~\eqref{E1} over the azimuth angle $\Phi_\re$, we have the number of photons emitted toward the photon sphere per unit time as 
\begin{align}
    d^5 \left( \frac{dN}{d\tau_\re} \right) 
    &= \frac{1}{2} \times 2\pi J^{(N)}_\re(\omega_\re, r_\re) \frac{\sqrt{f(r_\re)}}{r_\re} \frac{b}{\sqrt{(r_\re^2 / f(r_\re)) - b^2}} 
    \notag \\
    &\quad \times d b\, dV_\re d\omega_\re,
    \label{E2}
\end{align}
where we multiply a factor of $1/2$ taking into account that only photons with $0 \leq \Theta_\re \leq \pi/2$ can approach the photon sphere. 

The time element $d\tau_\re$ and volume element $dV_\re$ in a local inertial frame at $p_\re$ are respectively given by $d\tau_\re = \sqrt{f(r_\re)} \, dt$ and $dV_\re = (r_\re^2/\sqrt{f(r_\re)}) \sin \theta dr_\re d\theta d\phi$ in terms of the Schwarzschild coordinates. 
The integration of Eq.~\eqref{E2} over $\theta$ and $\phi$ leads to the number of photons emitted from a spherical shell of width $dr_\re$ with impact parameter $(b, b+db)$ per unit Schwarzschild time $t$ and unit frequency $\omega_\re$ as
\begin{align}
    d^2 \left(\frac{d^2N}{dt d\omega_\re} \right)
    = 4\pi^2 J^{(N)}_\re(\omega_\re, r_\re) \frac{b \,r_\re \sqrt{f(r_\re)}}{\sqrt{(r_\re^2 / f(r_\re)) - b^2}} dr_\re \, db.
    \label{E3}
\end{align}

Note that $\omega_\re$ above is given as the frequency in a local inertial frame at the emission point $p_\re$, i.e.,  
\begin{align}
    \omega_\re &= k^\mu (e^0)_\mu |_{p_\re} 
    \notag \\
    &= \frac{dt}{d\lambda} \sqrt{f(r)} \bigg|_{p_\re} 
    \notag \\
    &= \frac{E}{\sqrt{f(r_\re)}} ,
    \label{E4}
\end{align}
where $k^\mu = dx^\mu(\lambda) / d\lambda$ is the tangent vector of the geodesic of the photon with affine parameter $\lambda$, and $(e^0)_\mu$ is a member of the tetrad introduced in Eq.~\eqref{C5}.
In the last line, we have used Eq.~\eqref{B1}.
The frequency measured in a local inertial frame at the photon sphere located at $r= 3M$ is $\omega_\rc = E/ \sqrt{f(3M)} = \sqrt{3f(r_\re) } \,\omega_\re$.
Thus, the number of photons approaching the photon sphere per unit time $t$ and unit frequency $\omega_\rc$ with impact parameter $(b,b+db)$ is given by\footnote{Strictly speaking, the pericenter of a photon's trajectory depends on $b$ as Eq.~\eqref{B15}. Here, we approximate the radial coordinate of the trajectory as the value on the photon sphere, $r=3M$, for all photons.}
\begin{align}
    d \left(\frac{d^2N}{dt d\omega_\rc} \right)
    &= \frac{4\pi^2}{\sqrt{3}} 
    \int_{r_{\text{in}}}^{r_\text{out}} dr_\re \, J^{(N)}_\re \left( \frac{\omega_\rc}{\sqrt{3f(r_\re)}}, r_\re \right)
    \notag \\
    &\quad \times \frac{b\, r_\re}{\sqrt{ (r_\re^2 / f(r_\re)) - b^2 }}  \, db ,
    \label{E5}
\end{align}
where we assumed that the emission region is a sphere with inner diameter $r_\mathrm{in}$ and outer diameter $r_{\mathrm{out}}$.

\section{Light scattering by plasma}
\label{app:scattering}

In Sec.~\ref{sec:photon_sphere}, we have omitted the possibility that photons may be scattered by surrounding plasma during the propagation. Taking into account the finite mean free path of photons, the expression for the number of photons converted into axions per unit time and unit frequency \eqref{axion_rate_0-1} should be modified as 
\begin{align}
    \frac{d^2N_{\gamma \to a}}{dt d\omega_\rc}
    &= \int db \, \frac{1}{2} \left(\frac{d^3N}{dtd\omega_\rc db}\right) 
    P_{\gamma \to a} \left( \frac{T(b)}{\sqrt{3}} \right)
    \notag \\
    &\qquad \times \exp\left({-\frac{T(b)}{\sqrt{3}}\frac{1}{\ell}} \right)
    \label{axion_rate_sct_1}
\end{align}
where $T(b)$ is given by Eq.~\eqref{B17}, $P_{\gamma \to a}$ is given by Eq.~\eqref{axion_rate_0-2},  and $\ell$ is the mean free path of photons. 
As in Eq.~\eqref{axion_rate_0-3}, we replace $d^3N / dtd\omega_\rc db$ in the integrand by the value at $b = b_{\crit}$ for approximation. Then, we can perform the integration over $b \in (b_{\crit} , b_{\crit} + \sqrt{3} \epsilon^2 M /2)$ as 
\begin{align}
    \frac{d^2N_{\gamma \to a}}{dt d\omega_\rc}
    &\simeq \frac{1}{2} \left. \frac{d^3N}{dt d\omega_\rc db} \right|_{b = b_\crit}
    \notag \\
    &\quad \times 
    \left( \frac{\Delta_\rM}{\Delta_\osc / 2} \right)^2    
    \frac{\sqrt{3} \epsilon^2 M}{4}
    \frac{1}{1 + (3M/ \ell)}
    \notag \\
    &\quad \times \frac{(3M \Delta_\osc)^2}{(1 + (3M/\ell))^2 + (3M \Delta_\osc)^2 } .
    \label{axion_rate_sct_2}
\end{align}
It is obvious that the result in Eq.~\eqref{axion_rate_0-3} is reproduced when $3M / \ell \ll 1$. 
For photons with a frequency below the electron mass, the mean free path is given by 
\begin{align}
    \ell &= \frac{1}{\sigma_\mathrm{T} n_e} 
    \notag \\
    &= 1.5 \times 10^{24} ~ \si{cm} \left( \frac{\si{cm}^{-3}}{n_e} \right),
\end{align}
where $\sigma_\mathrm{T} = 8 \pi \alpha^2 / (3 m_e^2) = 0.67 \times 10^{-24} ~\si{cm}^{2}$ is the Thomson cross section, and $n_e$ is the electron density.
For black holes considered in this paper, the condition $3M / \ell \ll 1$ is satisfied. Thus, Eq.~\eqref{axion_rate_0-3} neglecting $\ell$ is applicable.

\bibliography{references}

\end{document}